\pdfoutput=1
 
\def\commentswitch#1{\iffalse#1\fi} 

\def\commentswitchsj#1{\iftrue#1\fi} 

\documentclass[a4paper,fleqn,usenatbib]{mnras}
\usepackage[T1]{fontenc}
\usepackage{ae,aecompl}

\usepackage{graphicx}	

\usepackage{amsmath}	
\usepackage{amssymb}	

\usepackage{dcolumn}     
\usepackage{bm}          
\usepackage{color}

\usepackage{lineno} 

\usepackage[dvipsnames]{xcolor}
\usepackage{todonotes,multirow,enumitem}
\usepackage{soul}
\usepackage[normalem]{ulem}
\usepackage{siunitx}

\usepackage{booktabs}

\usepackage{graphbox}

\usepackage{eso-pic}

\AddToShipoutPictureBG*{%
  \AtPageUpperLeft{%
    \hspace{0.75\paperwidth}%
    \raisebox{-3.5\baselineskip}{%
      \makebox[0pt][l]{\textnormal{DES-2020-0532}}
}}}%

\AddToShipoutPictureBG*{%
  \AtPageUpperLeft{%
    \hspace{0.75\paperwidth}%
    \raisebox{-4.5\baselineskip}{%
      \makebox[0pt][l]{\textnormal{FERMILAB-PUB-21-207-AE}}
}}}%

\newcommand{\ud}{\mathrm{d}}

\newcommand{\bea}{\begin{eqnarray}}
\newcommand{\eea}{\end{eqnarray}}
\newcommand{\beas}{\begin{eqnarray*}}
\newcommand{\eeas}{\end{eqnarray*}}  
\newcommand*{\swap}[2]{#2#1}

\newcommand{\hMpc}{h^{-1}{\rm\;Mpc}}

\newcommand{\degsqr}{{\rm\;deg^2}}

\newcommand\cosmosis{{\textsc{CosmoSIS}}}
\newcommand\cosmolike{{\textsc{cosmoLike}}}
\newcommand{\redmagic}{{\textsc{redMaGiC}}}
\newcommand\metacal{{\textsc{metacalibration}}}
\newcommand\mgcamb{{\tt MGCamb}}

\newcommand\multinest{{\tt multinest}}
\newcommand\halofit{{\tt halofit}}

\def\citepp#1{\citeauthor{#1} \citeyear{#1}}

\title[DMASS MG]{Probing gravity with the DES-CMASS sample and BOSS spectroscopy}

\author[DES Collaboration]{
\parbox{\textwidth}{
\Large
S.~Lee,$^{1}$
E.~M.~Huff,$^{2}$
A.~Choi,$^{3}$
J.~Elvin-Poole,$^{3,4}$
C.~ Hirata,$^{3,4}$
K.~Honscheid,$^{3,4}$
N.~MacCrann,$^{5}$
A.~J.~Ross,$^{3}$
M.~A.~Troxel,$^{1}$
T.~F.~Eifler,$^{6,2}$
% ...
\begin{center} (DES Collaboration) \end{center}
\emph{\normalsize Affiliations are listed at the end of the paper}
}}

\date{Accepted XXX. Received YYY; in original form ZZZ}

\pubyear{2021}

\begin{document}
\label{firstpage}
\pagerange{\pageref{firstpage}--\pageref{lastpage}}
\maketitle

\begin{abstract}
The DES-CMASS sample (DMASS) is designed to optimally combine the weak lensing measurements from the Dark Energy Survey (DES) and redshift-space distortions (RSD) probed by the CMASS galaxy sample from the Baryonic Oscillation Spectroscopic Survey (BOSS). In this paper, we demonstrate the feasibility of adopting DMASS as the equivalent of CMASS for a joint analysis of DES and BOSS in the framework of modified gravity. We utilize the angular clustering of the DMASS galaxies, cosmic shear of the DES \metacal\ sources, and cross-correlation of the two as data vectors. By jointly fitting the combination of the data with the RSD measurements from the CMASS sample and {\it Planck} data, we obtain the constraints on modified gravity parameters $\mu_0=-0.37^{+0.47}_{-0.45}$ and $\Sigma_0=0.078^{+0.078}_{-0.082}$. 
Our constraints of modified gravity with DMASS are tighter than those with the DES Year 1  \redmagic\ sample with the same external data sets by $29\%$ for $\mu_0$ and $21\%$ for $\Sigma_0$, and comparable to the published results of the DES Year 1 modified gravity analysis despite this work using fewer external data sets. This improvement is mainly because the galaxy bias parameter is shared and more tightly constrained by both CMASS and DMASS, effectively breaking the degeneracy between the galaxy bias and other cosmological parameters. Such an approach to optimally combine photometric and spectroscopic surveys using a photometric sample equivalent to a spectroscopic sample can be applied to combining future surveys having a limited overlap such as DESI and LSST. 
\end{abstract}



\section{Introduction}
\label{sec:intro}

Over the past three decades after the discovery of the accelerating Universe \citep{Riess1998,Perlmutter1999}, the $\Lambda$CDM model has been widely accepted as the simplest and the most successful concordance model in modern cosmology. 
This model is based upon a spatially-flat, expanding Universe governed by Einstein's General Relativity (GR) and whose components are dominated by roughly 25\% of cold dark matter (CDM) and 70\% of dark energy, which is commonly associated with a cosmological constant. The cosmological constant, denoted as $\Lambda$, can be cast in the model as a perfect fluid with the equation-of-state parameter of minus one in order to trigger the late-time cosmic acceleration. The $\Lambda$CDM model has been thoroughly validated through a broad range of stringent tests using cosmological data sets such as the Cosmic Microwave Background (CMB), Type Ia supernovae, baryon acoustic oscillation (BAO), the large-scale clustering of galaxies, and weak gravitational lensing. Despite the overall success of the $\Lambda$CDM model supported by many observations, however, several fundamental puzzles remain. One notable concern is that the cosmological constant has no explicit physical theory for its origin. In the context of quantum field theory, one may connect the cosmological constant with the vacuum energy associated with zero-level quantum fluctuations.  
However, this approach can be easily countered by $\sim 120$ orders of magnitude difference in the value of the vacuum energy predicted by quantum field theory and inferred from the LCDM model \citep{Weinberg1989}. Such discrepancy leads to the desire for alternative models beyond the $\Lambda$CDM cosmology.  

Modified gravity (MG) has been suggested as one of the strong candidates to explain the cosmic acceleration without introducing a cosmological constant. In such a theory, modified GR at cosmological scales naturally produces an acceleration identical to the one assumed in the $\Lambda$CDM model, without raising the same issues as a cosmological constant. In a phenomenological approach, the modification to GR is often parametrized as two MG parameters added to the gravitational potentials in the Friedmann equations. These MG parameters modify the growth equations derived from the Friedmann equations, and thereby any departure from GR appears as a change in the growth of structure and the deflection of light while keeping the same expansion history of the Universe as $\Lambda$CDM (see \citepp{Ishak2019} for an  overview of the theory and phenomenology of MG models). 
Hence, it is worth noting that probes sensitive to the growth of structure play a crucial role in testing deviations from GR. 

Redshift space distortions (RSD) and weak gravitational lensing have been used together as a popular combination of growth data to test GR \citep{Zhang2007, Reyes2010,Simpson2013,Blake2016,Torre2017,Ferte2019,Singh2019,DES-EXT,Planck2018CosmologicalParameter}. 
A general approach to combine these two probes is adding an independent measurement of $f(z) \sigma_8(z)$ from redshift-space distortions by a spectroscopic survey to a weak lensing measurement by a photometric survey. 
However, several papers \citep{Bernstein2011,Gaztanaga2012,Cai2012} have shown that a combined analysis of overlapping spectroscopic and weak lensing surveys can yield much stronger dark energy and growth constraints than a combination of independent RSD and weak lensing measurements.
%
The motivation for combining those two overlapping probes comes from the fact that RSD provides the constraints of the growth parameters only in combination with other parameters, i.e., the measurements of $f(z) \sigma_8(z)$ and $\beta=f(z)/b$ where $f(z)$ is the redshift dependent growth rate,  $\sigma_8(z)$ is the amplitude of the matter clustering at redshift $z$, and $b$ is galaxy bias. Galaxy bias has been a major source of uncertainty in  cosmological analyses of large-scale structure. 
In the MG framework, its impact is even more significant because one of the MG parameters modifying the Newtonian potential enters into $\sigma_8(z)$ and $f(z)$ through the growth factor term and is strongly degenerate with galaxy bias. 
Without any prior knowledge of galaxy bias, one cannot constrain $f(z)$ or $\sigma_8(z)$ independently, resulting in the degradation of the measurement of MG parameters\footnote{This statement about degeneracy assumes only linear scales, where MG can be modeled.}. 
Meanwhile, weak gravitational lensing directly measures the value of $\sigma_8$ today. The cross-correlation of RSD and weak lensing enables us to tighten the constraint of galaxy bias by breaking the $f$--$b$ degeneracy and allows a more precise inference of the underlying distribution of matter.


Several studies have taken the aforementioned approach to test GR by combining the RSD measurements from the Baryonic Oscillation Spectroscopic Survey \citep[BOSS;][]{Eisenstein2011BOSS} and weak lensing measurements from recent deep imaging surveys such as the Canada-France-Hawaii Telescope Lensing Survey \citep[CFHTLenS;][]{CFHTLenS}, Dark Energy Survey \citep[DES;][]{DESOverview}, Kilo-Degree Survey \citep[KiDS;][]{KIDS} and the Hyper Suprime-Cam Survey \citep[HSC;][]{HSCAihara}. The BOSS target galaxy samples, LOWZ and CMASS \citep{Reid2016}, are the largest galaxy spectroscopic samples yielding the best BAO and RSD measurements in the redshift range of $0.15 < z < 0.75$. The large sample size and the availability of spectroscopic redshifts have turned the galaxies in the samples into a popular candidate for gravitational lenses in the overlapping regions with the imaging surveys. By adopting BOSS galaxies as lenses, one can access better deep images while maintaining the strong constraining power of the galaxy clustering measurements from BOSS.

\cite{Alam2017TestingCMASS} tested gravity by combining galaxy-galaxy lensing from CFHTLenS \citep{Miyatake2015} with redshift space galaxy clustering from the BOSS CMASS sample \citep{Alam2015TheSDSS-III}. The galaxy-galaxy lensing signal was obtained around the CMASS galaxies living in the small overlapping area of $105 \degsqr$ between the BOSS and CFHTLenS footprints. \cite{Jullo2019} performed a similar analysis with the addition of a shape catalogue from CFHT-Stripe 82 \citep{CFHT-S82} to extend the available area for weak lensing to $250 \degsqr$. 
Unlike the previous cases, the KiDS survey was intentionally designed to mostly overlap with the BOSS and 2dFLenS surveys to maximize the number of reliable spectroscopic lenses in their full footprint \citep{KIDS}. \cite{KIDS450MG} and \cite{Amon2018MG} fully utilized the KiDS-450 footprint ($450 \deg^2$) to test gravity in a phenomenological approach using the BOSS galaxies as lenses on the KiDS imaging data. Later, \cite{KiDS1000MG} constrained the $f(R)$ gravity model with the BOSS galaxies over the KiDS-1000 footprint covering the increased area of $\sim 1,000 \degsqr$. As shown in these previous studies, however, the MG analyses performed with the spectroscopic samples have to face a limitation due to a fairly small overlapping region with spectroscopic surveys (mostly within a few hundreds of $\degsqr$), unless imaging surveys are planned in consideration of utilizing the existing spectroscopic information such as KiDS. 

The DES-CMASS galaxy sample from the Dark Energy Survey  (hereafter DMASS) has been developed to optimally combine the measurements of weak lensing from DES and RSD from BOSS by extending the available area for such analyses beyond the overlap between BOSS and DES \citep{DMASS}. 
The selection algorithm for DMASS is trained in the overlapping region ($123 \deg^2$) with galaxy colors and magnitudes, and then identifies CMASS-like galaxies in the rest of the DES footprint where the spectroscopic information is not available. 
The resulting DMASS sample replicates the statistical properties of the CMASS galaxy sample and populates the lower region of the DES Year 1 (Y1) wide-area survey footprint ($1,244 \deg^2$), excluding the overlap. 
Using the DMASS galaxies as lenses, one can obtain the measurements of galaxy clustering and galaxy-galaxy lensing equivalent to those that would have been measured if CMASS populated the full footprint of DES. These two clustering measurements, along with cosmic shear (the correlation of galaxy shapes) from the DES source galaxies, 
not only extract the full statistical power of DES, but they can also be efficiently combined with the measurement of redshift-space distortion from CMASS without introducing additional systematics parameters such as galaxy bias.

\begin{figure}
\centering
\includegraphics[width=0.49\textwidth]{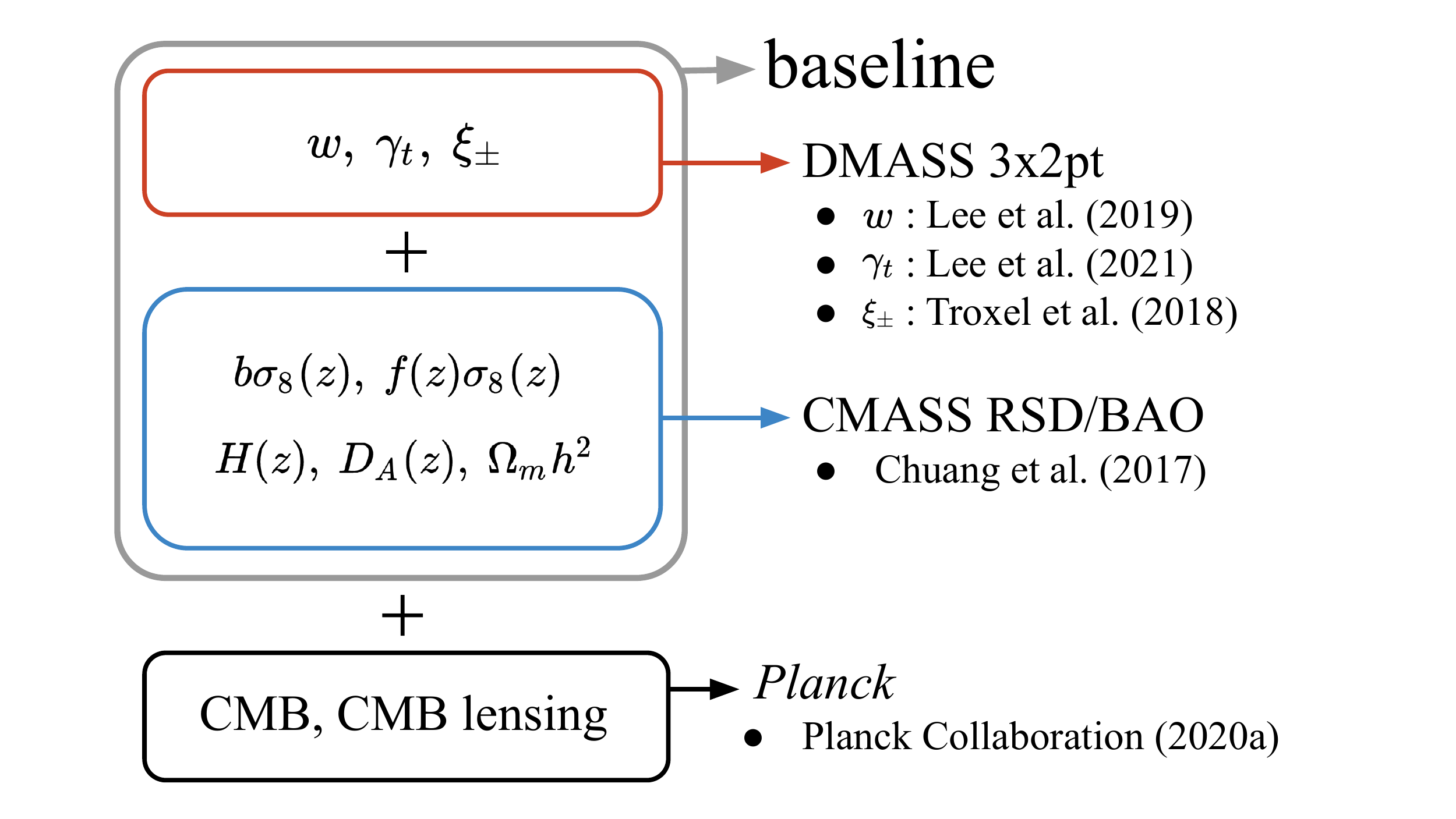}
\caption{ 
Summary of the different data sets in this analysis. The data in this paper consist of the DMASS 3x2pt measurements, the RSD and BAO measurements from the BOSS CMASS sample, and the CMB measurement from {\it Planck}.   
The DMASS 3x2pt measurement in the red box includes angular clustering ($w$), cosmic shear ($\xi_{\pm}$), the cross-correlation of the two (galaxy-galaxy lensing, $\gamma_t$) measured with the DMASS lenses and the DES Y1 \metacal\ sources. The measurement from CMASS provides five constraints on RSD and BAO in the blue box. See Section \ref{sec:data.baorsd} for a detailed description for these parameters. The combination of the red and blue boxes is defined as `baseline'. The galaxy bias parameter ($b$) is shared in this combination. The black box represents the CMB and CMB lensing measurements from {\it Planck} (see Section \ref{sec:data.cmb}). 
}
\label{fig:dataset}
\end{figure}

In this paper, we test the feasibility of this optimal combination method using the DMASS galaxy sample as lenses in the framework of phenomenological modified gravity. 
To isolate the results from any additional complexity arising from different probes, we adopt a minimal set of data as follows:  the combination of angular galaxy clustering, cosmic shear, and the cross-correlations of the two (galaxy-galaxy lensing) measured with the DMASS lenses and DES Y1 sources (hereafter DMASS 3x2pt), the RSD and  BAO measurements from the BOSS CMASS sample \citep{Chuang2017}, and {\it Planck} 2018 data \citep{Planck2018CosmologicalParameter}. 
The measurement of angular galaxy clustering used in this work is described in the original sample paper \citep{DMASS}. The measurement of galaxy-galaxy lensing and its calibration procedures are detailed in \citep{DMASS-GGL}. DES Y1 cosmic shear is adopted from \cite{Troxel2018}. 
Note that we define `DMASS 3x2pt + CMASS RSD/BAO' as a baseline as DMASS is designed to harness its maximum power in combination with CMASS. In Figure \ref{fig:dataset}, we summarize the data sets used in the analysis.  
We follow the methodology used in the DES Y1 analysis for modified gravity \citep[hereafter DESY1MG;][]{DES-EXT} and compare their MG constraints with ours to estimate the efficiency of this method. 

This paper is organized as follows. In the following section, we introduce the phenomenological parametrization of MG adopted in this paper. The data sets and theoretical predictions used to describe the data are detailed in Section \ref{sec:data}. 
In Section \ref{sec:analysis}, we describe our analysis methodology. 
In Section \ref{sec:robustness}, we present a series of validation tests for potential systematics that might bias the results. In Section \ref{sec:blinding}, we briefly describe how we blind the data to avoid confirmation bias. In Section \ref{sec:result}, we present our results and compare them with the results of DESY1MG. Our conclusions are presented in Section \ref{sec:conclusion}.

\section{Parametrization of modified gravity}

We parametrize departures from GR in a phenomenological way. This approach has an advantage in the sense that it does not require exact knowledge of the specific alternative theory but is still able to capture a generic deviation of the perturbation evolution from $\Lambda$CDM, by injecting two parameters into the perturbed Einstein's  equations (see \cite{Ishak2019} for a general overview of the phenomenological approach and its applications).

The perturbed Friedmann-Lema\^{i}tre-Robertson-Walker (FLRW) metric describing the $\Lambda$CDM cosmology is defined in terms of the two gravitational potentials $\Psi$ and $\Phi$ given as 
\bea
\ud s^2 = (1 + 2 \Psi) \ud t^2 - a^2(t) (1 - 2\Phi) \ud x^2~. 
\eea
The evolution of the two gravitational potentials are described by the two equations as follow:
\bea
k^2 \Phi &=& - 4\pi a^2 G \rho \Delta~, 
\label{eq:poisson1} \\
k^2 (\Psi - \Phi) &=& -12 \pi G a^2 (\rho + P )  \sigma~,
\label{eq:poisson2}
\eea
where $\Delta$ is the gauge-invariant density contrast, $\rho$ and $\sigma$ are the density and the shear stress, and $P$ is the pressure. 
For a negligible shear stress, the combination of the two equations leads to another set of the Poisson equation as follows:  
\bea
k^2 \Psi &=& - 4\pi G a^2 \rho \Delta~, \\
k^2 \Psi_{W} &=& -4 \pi G a^2 \rho \Delta ~.
\eea
where $\Psi_{W}$ is the Weyl potential defined as $\Psi_{W} = (\Psi + \Phi)/2$ which affects the propagation of light.  
The deviations from GR can be encapsulated in two parameters multiplied to these gravitational potentials as below: 
%
\bea
\Psi(k,a) &=& [1+\mu(k,a)]\Psi_{\rm GR}(k,a)~, \\
\Psi_W (k,a) &=& [1+\Sigma(k,a)]\Psi_{W, {\rm GR} }(k,a)~.
\eea
The gravitational acceleration of non-relativistic particles is determined by $\Psi$, and the paths of photons depend on $\Psi_{W}$.  Therefore, $\mu$ is sensitive to modifications to the structure growth, whereas $\Sigma$ governs modifications to the lensing of light. 
One can break the degeneracy between $\mu$ and $\Sigma$ by combining the measurements from galaxy clustering surveys with the measurements from weak lensing. 

In this paper, we adopt the time-evolving MG parameters following \cite{Ferreira2010,Simpson2013} given as
\bea
\mu(a) = \mu_0 \frac{\Omega_{\Lambda}(a)}{\Omega^0_{\Lambda}}, ~~~
\Sigma(a) = \Sigma_0 \frac{\Omega_{\Lambda}(a)}{\Omega^0_{\Lambda}}~,
\eea
where $\Omega_{\Lambda}^0 \equiv \Omega_{\Lambda}(a=1)$ is the dark energy density today so that $\mu_0$ and $\Sigma_0$ represent today's values of $\mu(1)$ and $\Sigma(1)$, respectively. Note that GR is restored for $\mu_0 = \Sigma_0 = 0$. 

\section{Data and measurements}
\label{sec:data}

In this section, we explain the data sets and measurements we use for the analysis. The primary data used in this work are the DMASS galaxy catalog \citep{DMASS} and \metacal\ shape catalog \citep*{Zuntz2018} from DES. Both catalogs are based on the images taken between Aug. 31, 2013 and Feb. 9, 2014 
during the first year observations of DES  \citep{DESCollaboration2006, Flaugher2015THECAMERA, DESCollaboration2017}. In Section \ref{sec:data.des}, we briefly describe the two catalogs and the 3x2pt measurements (galaxy clustering, tangential shear, and cosmic shear) obtained with these catalogs. 
We also utilize the RSD and BAO measurements extracted from the galaxy clustering of the BOSS CMASS galaxies \citep{Chuang2017}. Their measurements and corresponding covariance matrices are presented in Section \ref{sec:data.baorsd}. In Section \ref{sec:data.cmb}, we briefly describe the {\it Planck} CMB data we include to constrain the early universe. 

\subsection{Dark Energy Survey}
\label{sec:data.des}


\subsubsection{DMASS and \metacal\ catalogs}
\label{sec:data.catalogs}

\begin{figure}
\centering
\includegraphics[width=0.45\textwidth]{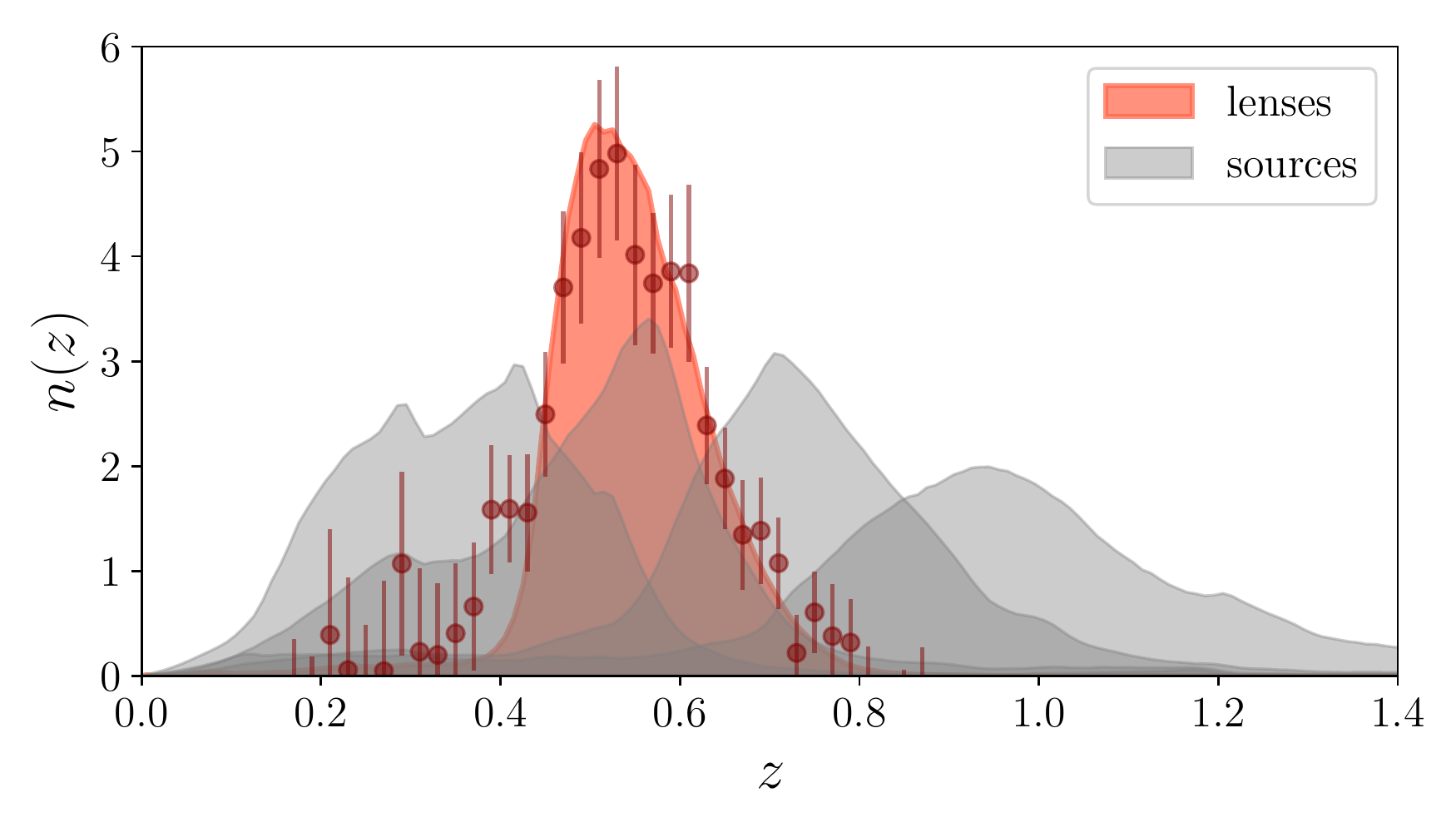}
\caption{ The spectroscopic redshift distribution of BOSS CMASS (red shaded) and the photometric redshift distributions of \metacal\ (grey shaded).
The maroon color points with error bars show the  redshift distribution of DMASS computed by cross-correlating the sample with the DES Y1 \redmagic\ sample. 
As the redshift distributions of CMASS and DMASS show good consistency, we use the redshift distribuion of CMASS as the true redshift distribution for a theoretical prediction for DMASS. 
The source sample, \metacal\, is divided into 4 tomographic bins ($0.2 < z_s < 0.43$, $0.43 < z_s < 0.63$,  $0.63 < z_s < 0.90$ and $0.90 < z_s < 1.30$) using the BPZ photometric redshift code. 
}
\label{fig:nz}
\end{figure}

\begin{figure*}
\centering
\includegraphics[width=0.22\textwidth]{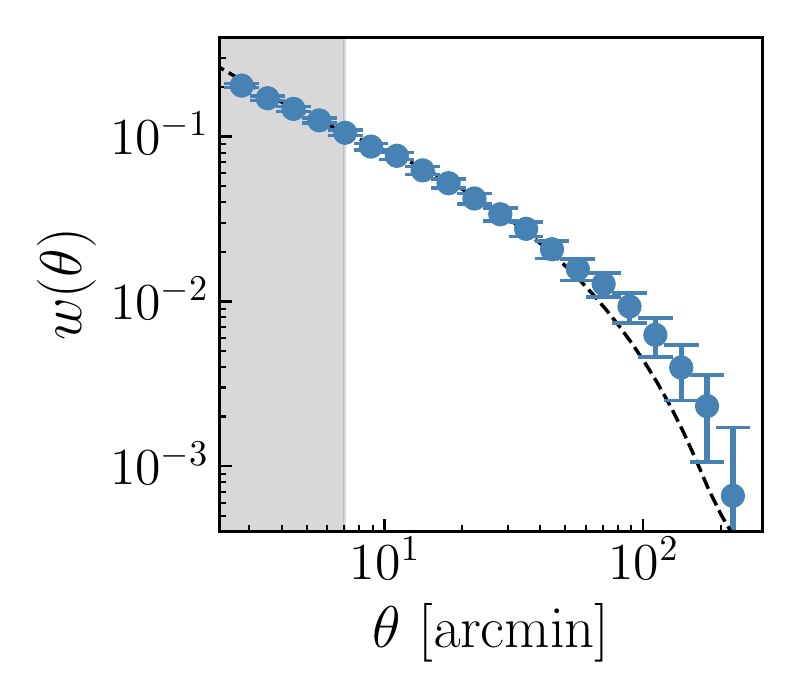}
\includegraphics[width=0.7\textwidth]{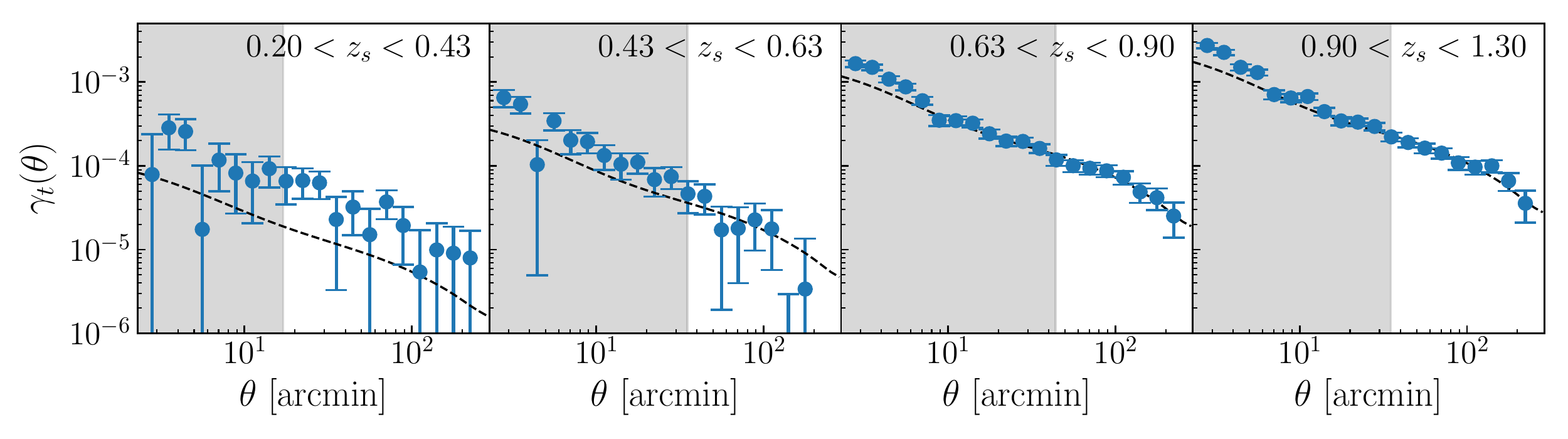}
\caption{The angular galaxy clustering (left) and tangential shear data (right) from \protect\cite{DMASS} and \protect\cite{DMASS-GGL}. The signals are measured with the DMASS (lenses) and DES Y1 \metacal\ (sources) catalogs. The dashed lines are the best-fitting MG predictions from the combination of the baseline case (DMASS 3x2pt + CMASS RSD/BAO) and {\it Planck} data. The shaded regions are discarded in the analysis to exclude the small scales where the phenomenological parametrization is not valid. The remaining data points are 16 points for $w(\theta)$ and 38 points for $\gamma_t(\theta)$.}
\label{fig:datavector1}
\end{figure*}

\begin{figure*}
\centering
\includegraphics[width=0.8\textwidth]{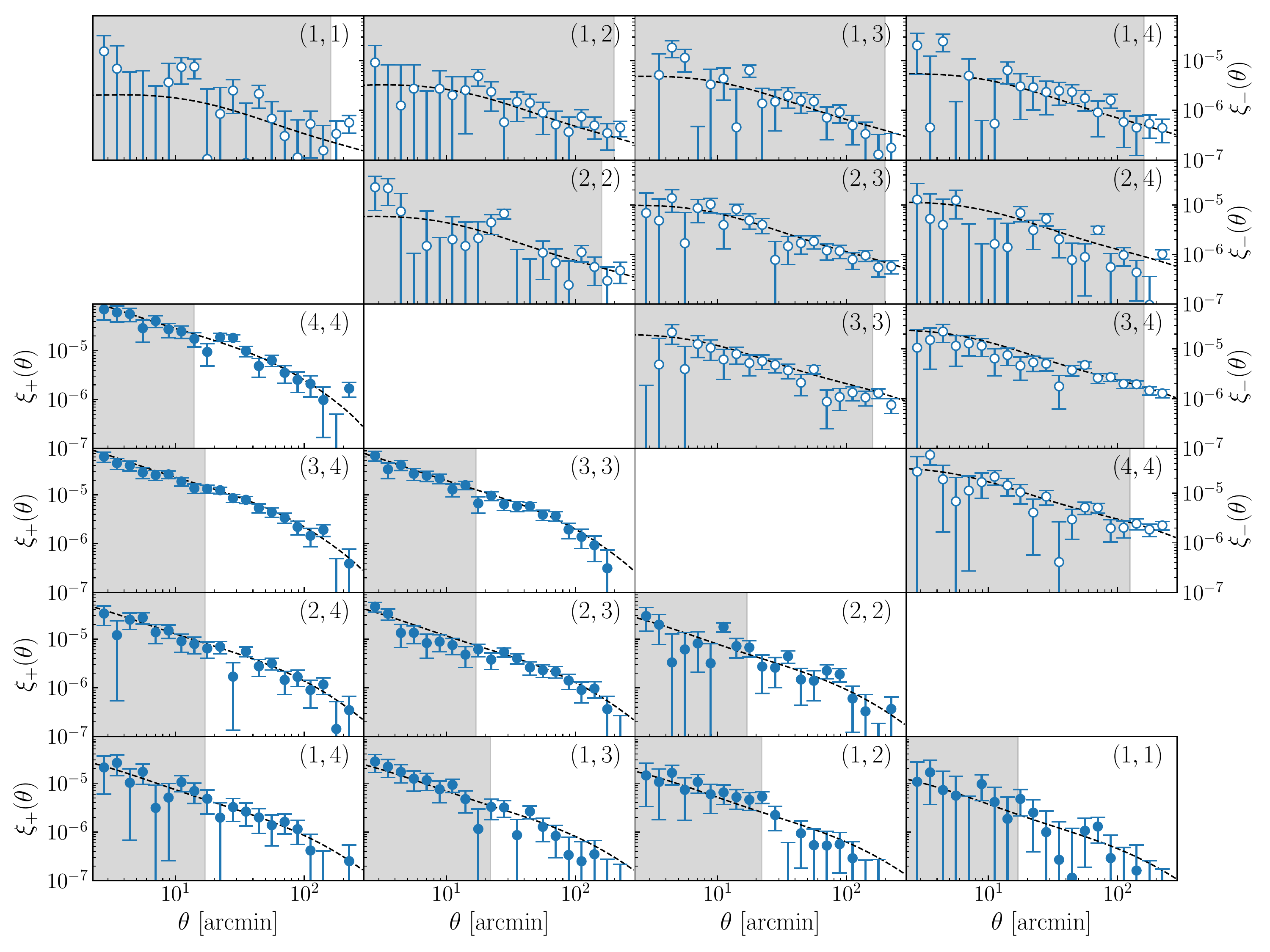}
\caption{ 
The cosmic shear data from \protect\cite{Troxel2018} measured with the DES Y1 \metacal\ catalog. The bottom triangle contains $\xi_{+}(\theta)$ and the top triangle contains $\xi_{-}(\theta)$. The dashed lines are the best-fitting MG predictions from the combination of the baseline case and {\it Planck} data. The labels ($i$, $j$) in the upper-right corner of each panel indicate the combination of the $i$th and $j$th source tomographic bins used to obtain each signal. The shaded regions are discarded for the cosmological analysis, leaving 119 points for $\xi_{+}(\theta)$ and 38 points for $\xi_{-}(\theta)$.} 
\label{fig:datavector2}
\end{figure*}

We utilize the DMASS galaxy sample  \citep{DMASS} as gravitational lenses. The DMASS sample is a photometric sample designed to replicate the statistical properties of the BOSS CMASS galaxy sample \citep{Reid2016}. 
The sample consists of $117,293$ effective galaxies selected from the DES  Y1  GOLD catalog \citep{Y1GOLD}, and covers the lower region of the DES Y1 wide-area survey footprint, excluding the overlapping area with BOSS near the celestial equator. The full coverage of DMASS is $1,244 \degsqr$ after masking.

%

The feasibility of using the DMASS sample as an equivalent of the CMASS sample has been studied in \cite{DMASS} and \cite{DMASS-GGL}. 
In \cite{DMASS}, the redshift distribution of DMASS was computed by cross-correlating the sample with the DES Y1 \redmagic\ galaxy sample \citep*{Rozo2016RedMaGiC:Data,ELVINPOOLE}.
Figure \ref{fig:nz} shows the redshift distributions of CMASS (red shaded region) and DMASS (maroon error bars). 
The two distributions show good agreement\footnote{The impact of the low- and high-ends and a bump at $z\sim0.4$ in the redshift distribution of DMASS has been tested in the appendices of \cite{DMASS} and \cite{DMASS-GGL}. Both have a negligible impact on the signals of galaxy clustering and tangential shear.}.
Therefore, we use the redshift distribution of CMASS as the true redshift distribution for a theoretical prediction for DMASS. 
In addition, \cite{DMASS} showed the consistency between DMASS and CMASS by comparing various statistical properties such as the number density, auto-angular clustering, and cross-angular clustering with the external surveys. 
The resulting values of the difference in the mean galaxy bias and mean redshift are $\Delta b = 0.044^{+0.044}_{-0.043}$ and $\Delta z = (3.51 ^{+4.93}_{-5.91}) \times 10^{-3}$, which indicate that the galaxy bias and mean redshift of two samples are consistent within $1\sigma$. 
For more description of the galaxy sample and sample selection algorithm, we refer readers to \cite{DMASS}. 

Source galaxies are selected from the DES \metacal\ catalog  \citep*{METACAL2,METACAL1,Zuntz2018}. 
Photo-$z$ of individual galaxies are evaluated by the Bayesian Photometric Redshift (BPZ) algorithm \citep*{BPZ,Hoyle2018}. 
As in \citet*{Zuntz2018,Prat2018DarkLensing} and \citet{Troxel2018}, sources are split into four redshift bins selected using BPZ: $0.2 < z < 0.43 $, $0.43 < z < 0.63 $, $0.63 < z < 0.90 $ and $0.90 < z < 1.30 $. 
The redshift distributions of the four source bins are plotted in Figure \ref{fig:nz} with the redshift distribution of lenses. 
The shear multiplicative biases, photo-$z$ biases, and their uncertainties related to this catalog are quantified in \citet*{Zuntz2018,Hoyle2018} and employed as priors in our analysis. See Section \ref{sec:analysis} for a detailed description. 

We do not split the lens sample in multiple redshift bins. Instead of using the five tomographic lens bins as done in DESY1MG, we consider only one lens bin along with the four source bins. 
This choice of one tomographic lens bin is motivated by the two reasons as follows: First, \cite{Chuang2017} split the LOWZ and CMASS sample into two bins (for a total of four bins) to increase the sensitivity of redshift evolution, but did not find improvement in terms of constraining different dark energy model parameters compared to the case of one bin. Second, splitting the DMASS sample into multiple redshift bins requires retraining the DMASS algorithm, and a series of validation tests performed in \cite{DMASS} should be followed. 
The combination of the one lens and four source bins results in one galaxy clustering signal, four galaxy-galaxy lensing signals, and twenty shear signals. In the following sections, we will describe the modeling and measurement of these two-point functions.

\subsubsection{Angular galaxy clustering}

The theoretical prediction for angular galaxy clustering is given as \citep{Kaiser1992,LoVerde2008}
\bea
w(\theta) = \frac{1}{2\pi} \int^{\infty}_{0} C_{gg} (\ell) J_{0} ( \ell \theta) \ell \ud \ell  ~,
\label{eq:wtheta}
\eea
where $J_n(x)$ is the $n$th order Bessel function of the first kind. 
The galaxy angular power spectrum $C_{gg}$ is the projection along the line of sight of the 3D power spectrum as given by
\bea
C_{gg} (\ell) = \int^{\infty}_{0} \ud \chi \frac{W_{g}^2 ( k, \chi)  }{\chi^2} P_{\delta\delta} (k, z(\chi))~,
\eea 
where $\ell$ denotes the angular multipole, $k=(\ell+1/2)/\chi$,  and $P_{\delta\delta}(k, z(\chi)) $ is the matter power spectrum. As the growth factor term $G^2(z)$ is contained in the matter power spectrum at $z$, modifications to gravity by $\mu$ enters into the matter power spectrum.  
The function $W_g(k, \chi)$ is the weight function for clustering defined as 
\bea
W_{g} (k,\chi) = b_g (k, z(\chi)) \frac{n_g (z(\chi)) }{\bar{n}_g} \frac{\ud z}{\ud \chi}~,
\eea
where $b_g$ is the galaxy bias of the lens galaxies. In this paper, we adopt the linear galaxy bias model as we restrict our analysis to sufficiently large scales.

\cite{DMASS} computed the angular clustering of DMASS to verify the consistency between DMASS and CMASS. By following the procedure described in that paper, we recompute the signal in the same manner, but with the number of angular bins increased from 10 to 20. 
The measurement of $w(\theta)$ is displayed in the first panel of Figure \ref{fig:datavector1} with the best-fitting MG prediction. 
As we obtained the same result except for the number of bins, we only briefly summarize the methodology below and refer readers to \cite{DMASS} for details.

The signal is evaluated in 20 logarithmically spaced angular bins over the range $2.5 \arcmin < \theta < 250 \arcmin$, using the Landy-Szalay estimator \citep{Landy1993BiasFunctions} as given by
\bea
w(\theta) = \frac{DD(\theta) - 2DR(\theta) + RR(\theta)}{RR(\theta)}~,
\eea
where DD, RR, and RR are the number of galaxy pairs, galaxy-random pairs, and random pairs separated by an angular distance $\theta$. 
Potential systematics that can bias the angular clustering were studied and corrected by applying weights to individual galaxies, as illustrated in Section 4 of \cite{DMASS}.
More details of the measurement procedures are described through Sections 4 and 5 in the same paper.

\subsubsection{Galaxy-galaxy lensing}
\label{sec:data.ggl}

We use the tangential shear as an observable for galaxy-galaxy lensing. The tangential shear correlating with the $i$th source bin is obtained from the Fourier transform of the angular power spectrum as follows:
\bea
\gamma^{i}_{t} (\theta) = \frac{1}{2\pi} \int^{\infty}_{0} C^{i}_{g \kappa} (\ell) J_{2} (\ell \theta) \ell \ud \ell  ~,
\eea
where $J_n(x)$ is the $n$th order Bessel function of the first kind. 
The angular power spectrum for galaxy-galaxy lensing  correlating with $i$th source bin takes the form
\bea
C^{i}_{g \kappa} (\ell) = \int^{\infty}_{0} \ud \chi \frac{W_{g} ( k, \chi) W^i_{\kappa} (\chi) }{\chi^2} P_{\delta\delta} (k, z(\chi))~.
\eea
The integral along the line of sight indicates that weak lensing radially projects the density fluctuations between us and the source galaxies. 
The function $W(\chi)$ is the geometric weight function describing the lensing efficiency defined as
\begin{flalign}
&~~W^i_{\kappa}(\chi) = \frac{3 H_0^2 \Omega_m}{2c^2}  \frac{\chi}{a(\chi)}  \\
&~~~~~~~~ \times \int^{\infty}_{\chi} \ud \chi' \frac{n^i_{\kappa}(z(\chi') ) \ud z/\ud \chi' }{ \bar{n}^i_{\kappa} } \frac{\chi' - \chi}{\chi}[1 + \Sigma(\chi')]~  
\label{eq:w_kappa}
\end{flalign}
in terms of the source distribution $n_{\kappa}(\chi)$. Modifications to GR in lensing appear as $[1 + \Sigma(\chi)]$. Modifications by $\mu$ enters in the matter power spectrum as shown in the case of $w(\theta)$. Thus, galaxy-galaxy lensing is sensitive to both modifications by matter and relativistic particles.  

We utilize the measurement of tangential shear around the DMASS galaxies presented in \cite{DMASS-GGL} as a data vector for $\gamma_t$ in the cosmological analysis. 
With the lens and four source bins, the mean tangential shear is computed by averaging over lens-source pairs as below:
\bea
\gamma_t(\theta) = \langle \gamma_{+} (\theta) \rangle =\frac{1}{\langle R \rangle } \frac{\sum_{j} w_{ls,j} \gamma_{+,j} }{\sum_{j} w_{ls,j} }~.
\label{eq:gamma}
\eea
The notation $w_{ls}$ is a combination of weights associated with each lens-source pair.
The value $\langle R \rangle$ in the denominator is the mean shear response averaged over the sources, which is defined as the sum of the measured shear response ($R_{\gamma}$) and shear selection bias  correction matrix ($R_S$) for \metacal ~as follows: 
$\langle R \rangle = \langle R_{\gamma} \rangle + \langle R_S \rangle$.
To ensure the measured signal is free from various systematic effects, \cite{DMASS-GGL} performed tests for the mean cross-component of the shear and estimated the impact of observing conditions. The signals were also corrected for the boost factor. 
The cross-correlation coefficient between the amplitude of the galaxy clustering of CMASS and the tangential shear of DMASS was found to be consistent with unity down to the scale of $4\hMpc$. 
The measurement is shown in Figure \ref{fig:datavector1} with the best-fitting MG predictions from the combination of the baseline  and {\it Planck} data. 
In Figure \ref{fig:datavector1}, the tangential shear signal from the first source bin  ($0.20 < z < 0.43$) is higher than predicted in theory. The same tendency is shown in comparison with the $\Lambda$CDM prediction. \cite{DMASS-GGL} suggests that the interplay between an unmodeled local peak in galaxy bias of DMASS at low redshifts and the first source bin partially overlapping with the low-redshift end of the DMASS tomographic bin causes the discrepancy. 
%
%
We will discuss more details about this discrepancy and its impact on modified gravity constraints in Section \ref{sec:result}. 
For more description of the tangential shear measurement, we refer readers to \cite{DMASS-GGL}.

\subsubsection{Cosmic shear}

The angular correlation function for cosmic shear correlating the source redshift bins $i$ and $j$ is expressed in terms of the cosmic shear power spectrum $C_{\kappa\kappa}$ given as 
\bea
\xi^{i,j}_{\pm} (\theta)= \frac{1}{2\pi} \int^{\infty}_{0} C_{\kappa\kappa}^{i,j}(\ell) J_{0/4} (\ell \theta) \ell \ud \ell ~,
\eea
where $J_n(x)$ is the $n$th order Bessel function of the first kind. 
The cosmic shear power spectrum is defined with the two lensing kernels as
\bea
C_{\kappa\kappa}^{i,j} (\ell) = \int^{\infty}_0 \ud \chi \frac{ W^i_{\kappa} (\chi) W^j_{\kappa} (\chi)}{\chi^2} P_{\delta\delta} ( k,  z(\chi) )~. 
\eea
As shown in the case of tangential shear, both $\mu$ and $\Sigma$ affect the signal but modifications by $\Sigma$ is dominant due to the double lensing kernels. 

We utilize the shear measurement from \cite{Troxel2018} obtained with the DES \metacal\ galaxies.  \cite{Troxel2018} presents the auto- and cross-correlation functions $\xi_{\pm}^{ij}(\theta)$ of the source galaxies divided into four redshift bins over scales $2.5 \arcmin < \theta < 250 \arcmin$. 
Systematics related to the source catalog are parametrized in the same manner as described in \cite{Troxel2018}. 
Figure \ref{fig:datavector2} shows the measurements of $\xi_{\pm}(\theta)$. The bottom triangle contains $\xi_{+}(\theta)$ and the top triangle contains $\xi_{-}(\theta)$. The dashed lines are the best-fitting MG predictions from the combination of the DMASS 3x2pt, CMASS RSD/BAO, and {\it Planck} data. The labels ($i$, $j$) in the upper-right corner of each panel indicate the combination of the $i$th and $j$th source tomographic bins used to obtain each signal. The lowest index indicates the lowest tomographic bin. See \cite{Troxel2018} for a detailed description.

\subsubsection{Angular scale cuts}
\label{sec:data.scalecut}

The phenomenological approach such as the $\mu,\Sigma$ parametrization is based on the linearly perturbed Einstein equations, so the scales should be limited to the linear regime only. To deal with the nonlinear scales  accurately, using either semi-analytical methods or N-body simulation based on a specific MG model is necessary. 

Following the {\it Planck} 2018 analysis \citep{Planck2018CosmologicalParameter}, we restrict ourselves to observables sensitive to the linear scales only. We compute the difference between the nonlinear and linear-theory predictions in the standard LCDM model in our fiducial cosmology as $\Delta \chi^2 = (d_{\rm nl} - d_{\rm lin})^T C^{-1} (d_{\rm nl} - d_{\rm lin})$. 
The nonlinear predictions are obtained using  \halofit\ \citep{Takahashi2012} implemented in \cosmosis.
Then, we identify the single data point that contributes most to this quantity, and remove that data point, iterating until the quantity reaches $\Delta \chi^2 < 1$. 
The shaded regions in Figures \ref{fig:datavector1} and \ref{fig:datavector2} indicate the data points removed. Through this process, we obtain 119 data points for $\xi_+(\theta)$, 18 points for $\xi_-(\theta)$, 38 points for $\gamma_t(\theta)$, and 16 points for $w(\theta)$.

\subsubsection{Covariance}
\label{sec:data.cov}

Our measurements of $w(\theta)$, $\gamma_t(\theta)$ and $\xi_{\pm}(\theta)$ are correlated across angular and source redshift bins. The joint covariance for these  measurements is computed by \cosmolike\ \citep{COSMOLIKE} using the halo-based approach \citep{Cooray2002}, assuming a $\Lambda$CDM cosmology. 
The covariance is calculated as the sum of Gaussian covariance, non-Gaussian covariance, and the super-sample covariance as described in \citet*{Krause2017}. 

We assume there is no cross-correlation between surveys as the DES and BOSS areas used do not overlap\footnote{The DMASS sample does not cover the overlapping area between BOSS and DES since the sources from the area were used to train the DMASS algorithm. Thereby, the angular clustering and tangential shear signals used in this work are computed excluding the overlap. The shear measurement from \cite{Troxel2018} did not use the overlapping area as well.}, and the CMB data is most sensitive to higher redshift than the galaxy surveys are. The galaxy surveys also form a small fraction of the full-sky CMB measurements.

\subsection{RSD and BAO measurements from BOSS CMASS}
\label{sec:data.baorsd}

Redshift space distortions (RSDs) are one of the most promising probes for testing gravity. 
On large scales, peculiar velocities of galaxies follow infall of matter towards high-density regions \citep{Kaiser1984}, and through that, they are sensitive to the growth rate of structure. 

Under the assumption of linear theory, 
the galaxy power spectrum in the redshift space ($P^s_g$) can be related to the real space matter power spectrum ($P_m^r$) by
\bea
P^s_g(k,\mu,z) = b(z) [ 1+ \beta(z) \mu^2 ]^2 P_m^r (k,z)~,
\label{eq:rsd}
\eea
where $b(z)$ is the galaxy bias, $\beta(z)$ denotes the amplitude of the RSD defined as  $\beta \equiv f(z)/b(z)$,  
$f(z)$ is the structure growth rate defined as $f\equiv d\ln D/d\ln a$ in terms of the growth factor $D(a)$, $\mu$ is the cosine of the angle with the line of sight. As $\sigma_8^2$ is the amplitude of the matter power spectrum, Equation \eqref{eq:rsd} implies that RSD probes $b(z) \sigma_8 (z)$ and $f(z) \sigma_8 (z)$. Modifications to GR emerge in the term $f(z)\sigma_8(z)$ and also in the growth factor $D(z)^2$ in the matter power spectrum.

The measurements of RSD that we use in this study are derived from the Baryon Oscillation Spectroscopic Survey \citep[BOSS;][]{Eisenstein2011BOSS, Bolton2012SpectralSurvey, Dawson2013TheSDSS-III}.
BOSS targeted two distinct samples known as LOWZ at $0.15 < z < 0.43$ and CMASS at $0.43 < z < 0.75$ \citep{Reid2016}. The higher redshift sample, CMASS, was designed to select a stellar mass-limited sample of objects of all intrinsic colors, with a color cut that selects almost exclusively on redshift \citep{Reid2016}. The DMASS sample we use as lenses is designed to mimic this sample.
\cite{Chuang2017} present cosmological constraints from galaxy clustering of the LOWZ and CMASS samples.
They provide the growth rate and mean galaxy bias combined with the amplitude of mass fluctuation ($f(z) \sigma_8(z)$ and $b(z) \sigma_8(z)$) at different redshift points\footnote{As we do not have a pipeline to analyze the BOSS measurement of the two-point function directly, we had to rely on the cosmological inferences instead. \cite{Chuang2017} is the only published measurement of the BOSS CMASS galaxy clustering that provides the constraint on $b\sigma_8(z)$ and its covariance with other cosmological parameters. Future analyses using the same strategy will likely need to consider the full analysis of the two-point function to make use of  the BAO reconstruction that is not included in \cite{Chuang2017}.}. In this work, we utilize their results measured at the mean redshift of CMASS ($z=0.59$). The galaxy bias parameter $b$ here is shared with the galaxy bias incorporated in the angular galaxy clustering and tangential shear of DMASS. 


We also utilize the BAO measurements from the same work  to constrain the geometry of the background universe. The BAO measurements chosen are the constraints of the Hubble parameter ($H(z)$), the comoving angular diameter distance ($d_A(z)$) at the mean redshift of CMASS ($z=0.59$), and the matter density fraction ($\Omega_m h^2$).

We use the full covariance matrix between those RSD and BAO parameters given in the same work.

\subsection{CMB \& CMB lensing}
\label{sec:data.cmb}

Cosmic Microwave Background (CMB) is one of the most powerful cosmological probes. Through its  primary anisotropies such as temperature and polarization power spectra, the CMB tightly constrains the background geometry and the initial conditions. The secondary anisotropies such as the Integrated Sachs-Wolfe-Effect (ISW) and the CMB lensing at late times are more relevant to scalar mode perturbations and the growth of large-scale structure. 

The ISW effect is caused by time variations in the gravitational potentials \citep{ISW}. When CMB photons travel from the surface of the last scattering to us, they gain energy while falling down gravitational potential wells, but they lose it again while climbing out of the potential wells. However, dark energy components or modifications to gravity can cause stretching in the potential well, resulting  in a net gain in energy for the photons during their journey through the potential wells.

The resulting temperature perturbation is given by 
\bea
\frac{\delta T}{T} (\hat{n}) = - \int^{\eta_*}_{\eta_0} \ud \eta \frac{\partial (\Psi + \Phi)}{\partial \eta}~,
\eea
where $T$ is the CMB temperature, $\eta$ is the conformal time defined as $\ud \eta \equiv \ud t /a$,  $\eta_*$ is the conformal time at the surface of last scattering and $\eta_0$ at the observer. Note that $(\Psi +\Phi)$ in the integral can be expressed in terms of the Weyl potential as $2 \Psi_W$. 
By changing the gravitational potentials and their time evolution (growth), modifications to GR affect the ISW and modify the amplitude of the CMB temperature power spectra at the largest angular scales ($\ell < 10$).

CMB lensing refers to the deflections of CMB photons by the intervening matter while traveling from the surface of last scattering to us. Hence, CMB lensing is sensitive to the distribution and growth rate of large-scale structures and their associated gravitational potential. The deflections of CMB photons smear out the CMB temperature power spectra and also induce non-Gaussianities in the temperature and polarization maps \citep{Bernardeau1998, Zaldarriaga1998,Okamoto2003}.

\begin{table*}
\caption{
Parameters and priors used to describe the measured galaxy-galaxy lensing signal. `Flat' is a flat prior in the range given while `Gauss' is a Gaussian prior with mean $\mu$ and width $\sigma$. Priors for the tomographic shear and photo-z bias parameters $m^i$ and $\Delta z_{\rm src}^i$ are identical to DESY1MG.
}
\label{tab:params}
\resizebox{\textwidth}{!}{%
\begin{tabular}{@{}llllllccccccccclc@{}}
\toprule
 &  & \multicolumn{1}{c}{Parameter}          &  &  &  & Notation               &  &  &  & Fiducial               &  &  &  & Prior                                     &  &                      \\ \midrule
 &  & Normalized matter density              &  &  &  & $\Omega_m$             &  &  &  & 0.306                  &  &  &  & Flat (0.1, 0.9)                           &  &                      \\
 &  & Normalized baryon density              &  &  &  & $\Omega_b$             &  &  &  & $4.845 \times 10^{-2}$ &  &  &  & Flat (0.03, 0.07)                         &  & \multicolumn{1}{l}{} \\
 &  & Amplitude of primordial power spectrum &  &  &  & $A_s$                  &  &  &  & $2.196 \times 10^{-9}$ &  &  &  & Flat ($5\times 10^{-10},5\times 10^{-9}$) &  &                      \\
 &  & Power spectrum tilt                    &  &  &  & $n_s$                  &  &  &  & 0.968                  &  &  &  & Flat (0.87, 1.07)                         &  &                      \\
 &  & Hubble parameter ($H_0 = 100h$)        &  &  &  & $h$                    &  &  &  & 0.678                  &  &  &  & Flat (0.55, 0.90)                         &  &                      \\
 &  & Normalized neutrino density            &  &  &  & $\Omega_{\nu}$         &  &  &  & $6.50 \times 10^{-4}$  &  &  &  & Flat ($5\times 10^{-4}$,$10^{-2}$)        &  &                      \\
 &  & Optical depth                          &  &  &  & $\tau$                 &  &  &  & 0.081                  &  &  &  & Flat (0.01, 0.20)                         &  &                      \\ \midrule
 &  & Modified gravity parameter             &  &  &  & $\mu_0$                &  &  &  & 0.0                    &  &  &  & Flat (-3.0, 3.0)                          &  & \multicolumn{1}{l}{} \\
 &  & Modified gravity parameter             &  &  &  & $\Sigma_0$             &  &  &  & 0.0                    &  &  &  & Flat (-3.0, 3.0)                          &  & \multicolumn{1}{l}{} \\ \midrule
 &  & Linear galaxy bias (lens)              &  &  &  & $b$                  &  &  &  & 2.0                    &  &  &  & Flat (0.8, 3.0)                           &  &                      \\
 &  & Intrinsic alignment amplitude                   &  &  &  & $A_{\rm IA}$           &  &  &  & 0.0                    &  &  &  & Flat (-5.0, 5.0)                          &  &                      \\
 &  & Intrinsic alignment scaling                   &  &  &  & $\eta_{\rm IA}$        &  &  &  & 0.0                    &  &  &  & Flat (-5.0, 5.0)                          &  &                      \\
 &  & Lens redshift bias                     &  &  &  & $\Delta z_{\rm lens}$  &  &  &  & 0.0035                    &  &  &  & Gauss ( 0.0035, 0.005)                       &  &                      \\
 &  & Source photo-z bias ($i=1$)              &  &  &  & $\Delta z^1_{\rm src}$ &  &  &  & -0.001                 &  &  &  & Gauss (-0.001, 0.016)                     &  &                      \\
 &  & Source photo-z bias ($i=2$)              &  &  &  & $\Delta z^2_{\rm src}$ &  &  &  & -0.009                 &  &  &  & Gauss (-0.009, 0.013)                     &  &                      \\
 &  & Source photo-z bias ($i=3$)              &  &  &  & $\Delta z^3_{\rm src}$ &  &  &  & 0.009                  &  &  &  & Gauss (0.009, 0.011)                      &  &                      \\
 &  & Source photo-z bias ($i=4$)              &  &  &  & $\Delta z^4_{\rm src}$ &  &  &  & -0.018                 &  &  &  & Gauss (-0.018, 0.022)                     &  &                      \\
 &  & Shear calibration bias $(i\in\{1,2,3,4\})$                &  &  &  & $m^i$                  &  &  &  & 0.012                  &  &  &  & Gauss (0.012, 0.023)                      &  &                      \\ \bottomrule
\end{tabular}%
}
\end{table*}

In this work, we utilize the {\it Planck} 2018 likelihood described in \cite{Planck2018CosmologicalParameter}. 
We use the {\it Planck} temperature and polarization auto- and cross-multipole power spectra denoted as $C^{TT}_{\ell}, C^{TE}_{\ell}, C^{EE}_{\ell}$, and $C^{BB}_{\ell}$. Specifically, we use the full range of $C^{TT}_{\ell},C^{TE}_{\ell}, C^{EE}_{\ell}$ from $29 < \ell < 2509$ and the low-$\ell$ polarization data including $C^{BB}_{\ell}$ from $2<\ell<29$. 
We also make use of the {\it Planck} CMB lensing likelihood from temperature only \citep{Planck2018Lensing}.

\section{Analysis}
\label{sec:analysis}

In this section, we describe our methodology of measuring the cosmological constraints in the framework of modified gravity. 

To compute a theoretical prediction, we use \cosmosis\footnote{\url{https://bitbucket.org/joezuntz/cosmosis}} \citep{COSMOSIS} with a version of \mgcamb\footnote{\url{https://github.com/sfu-cosmo/MGCAMB}} \citep{mgcamb1, mgcamb2, mgcamb3, camb} modified to include the $\Sigma$ and $\mu$ parametrization. We utilize the  \halofit\ prescription \citep{Takahashi2012} to compute the nonlinear power spectrum. 


We perform Markov Chain Monte-Carlo likelihood analyses using the \multinest\ algorithm \citep{multinest1,multinest2,multinest3} implemented in the \cosmosis\ package. 
Our analysis spans the parameter set $\{ \Omega_m, h_0, \Omega_b,  n_s, A_s, (\tau, \Omega_{\nu}) \}$ where the parentheses around the optical depth parameter and the neutrino density indicate that they are used only in the analysis combinations that use the CMB data. In addition to this set of $\Lambda$CDM parameters, we vary $\{\mu_0, \Sigma_0 \}$ to test gravity. 
Along with the parameter set, we also vary nuisance parameters describing the photo-z and shear systematics for different tomographic bins, and model parameters for the intrinsic alignment. Since we use the identical source sample as DESY1MG, we adopt the same models for the shear and photo-$z$ systematics \citep*{Krause2017}. 
The complete set of varied parameters and priors is summarized in Table \ref{tab:params}. The priors imposed in this analysis are identical to those of DESY1MG except for the lens galaxy bias ($b$) and lens redshift bias ($\Delta z_{\rm lens}$), summarized below.

\begin{enumerate}[leftmargin=*, label=(\roman*)]
  \itemsep1em
  \item Lens galaxy bias ($b$): the linear galaxy bias parameter for lenses. Since we select fairly large scales where the linear assumption is valid, we use a constant galaxy bias over the redshift range. The redshift evolution of CMASS and DMASS is negligible, which is illustrated in \cite{Salazar-Albornoz2017} and \cite{DMASS}. The scale-independence of the DMASS galaxy bias over the scales of interest is shown in \cite{DMASS-GGL}. The galaxy bias parameter is shared with the galaxy bias in $b\sigma_8$ from the BOSS measurements described in Section \ref{sec:data.baorsd}. 
  \item Lens redshift bias ($\Delta z_{\rm lens}$): we model the redshift distribution of lenses as $n_{\rm true}(z) = \hat{n}(z - \Delta z_{\rm lens})$, where $\hat{n}$ is the measured redshift distribution, and $\Delta z_{\rm lens}$ is a redshift bias parameter. 
\cite{DMASS} constrained $\Delta z_{\rm lens}$ by jointly fitting the residuals of the angular correlation and clustering-z measurements, and obtained  $\Delta z = 3.5 \times 10^{-4}$ with its uncertainty of $\sigma_{\Delta z} = 0.5 \times 10^{-3}$. We take those values as the mean and width of a Gaussian prior, respectively.  
\end{enumerate}

The likelihood of the combined probes is evaluated by the sum of individual log-likelihoods given as  
\bea
\ln \mathcal{L}(p) = - \frac{1}{2} \sum_i  \chi^2_{i}(p)~,
\eea
where $p$ is the set of varied parameters, the subscript `$i$' denotes the $i$th data set. The value of $\chi^2$ is estimated as below:
\bea
\chi^2(p) = \sum_{j,k} ( {\bf{d}} - {\bf{d}}_{\rm th}(p)  ) _j {\rm {\bf{C}}}^{-1}_{jk} ( {\bf{d}} - { \bf{d}}_{\rm th}(p)  )_k^T ~,
\eea
where $\bf{d}_{\rm th}$ and $\bf{d}$ are theoretical and measured datavectors, and $\bf{C}$ is the covariance matrix. We assume there is no cross-correlation between surveys as stated in Section \ref{sec:data.cov}. 

\section{Robustness tests}
\label{sec:robustness}

\begin{figure*}
\centering
\includegraphics[width=0.8\textwidth]{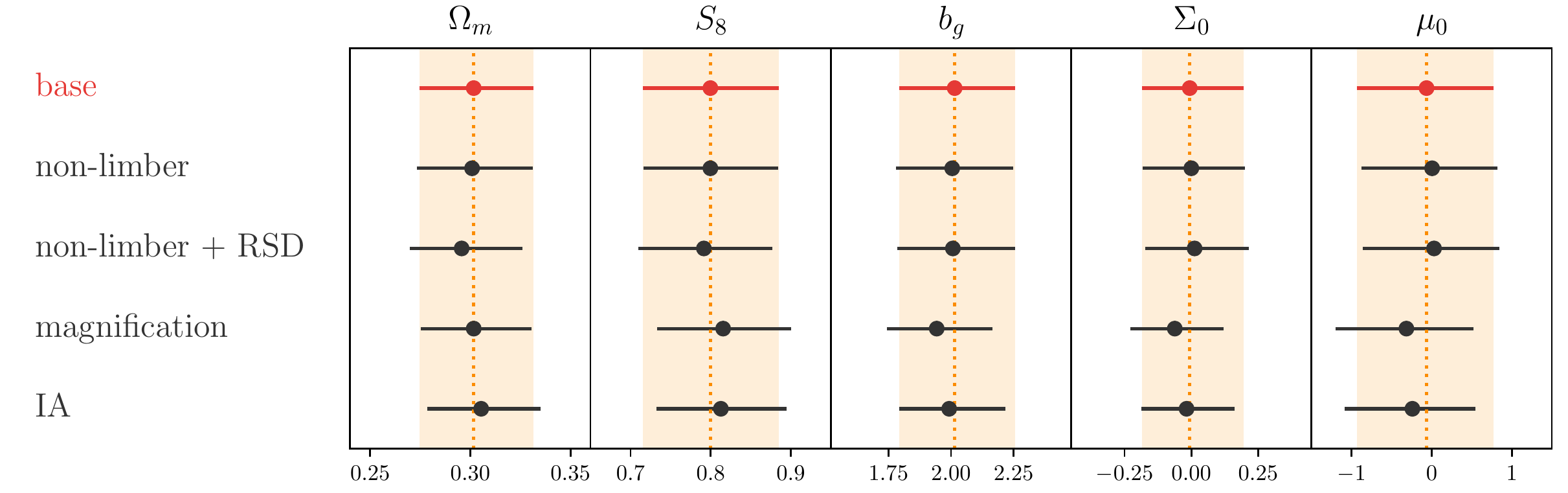} \\
$~$ \\
\includegraphics[width=0.8\textwidth]{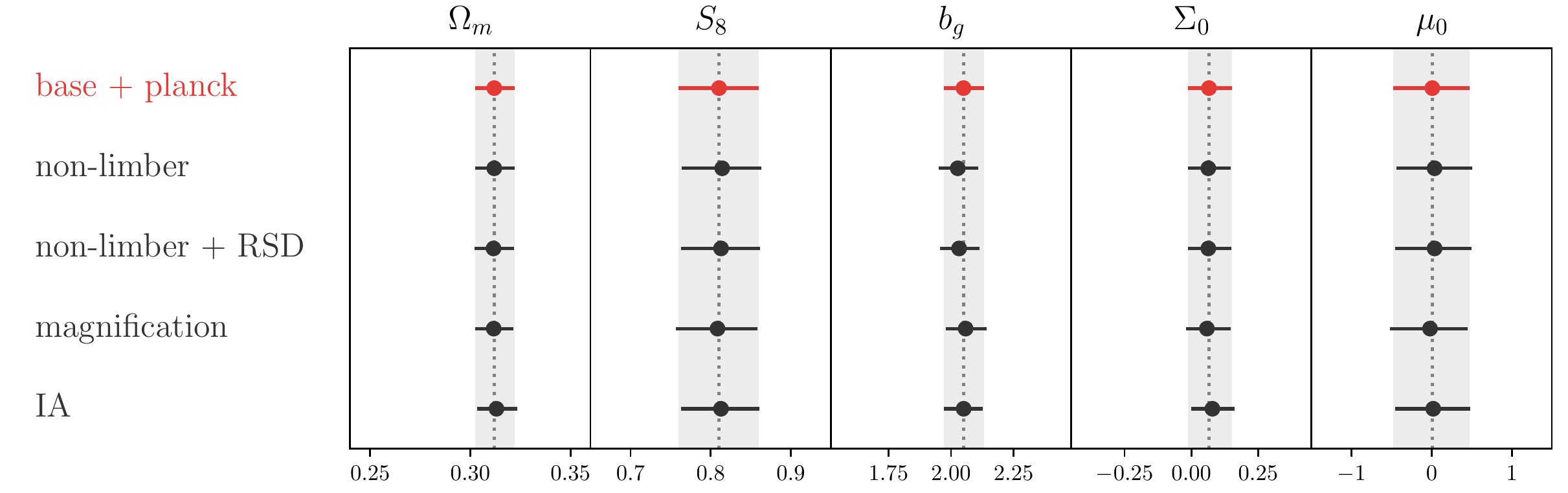}
\caption{Marginalized 1D posterior constraints on ${\Omega_m, S_8, b, \Sigma_0, \mu_0}$ obtained from DMASS 3x2pt + CMASS RSD/BAO (baseline; top) and DMASS 3x2pt + CMASS RSD/BAO + {\it Planck} (bottom). The different rows correspond to the different systematics injected in each of the simulated data vectors. The systematics tested here are detailed in Section \ref{sec:robustness}. These data vectors are analyzed by the fiducial pipeline to test the impact of systematics associated with each data vector on the parameter constraints.   
For all cases, the posteriors for these cosmological parameters remain well within $1\sigma$ of the fiducial constraints. 
}
\label{fig:sys-mg}
\end{figure*}

In this section, we perform a series of tests to show the robustness of our results to modeling assumptions and approximations. We thoroughly follow the procedures illustrated in DESY1MG. First, we generate a set of simulated data vectors of $w(\theta)$, $\gamma_t(\theta)$, and $\xi_{\pm}(\theta)$ shifted with the addition of a systematic effect that is not included in our analysis pipeline. Then, each of the simulated data vectors is individually analyzed with our fiducial pipeline by using the methodology described in Section \ref{sec:analysis}. Finally, we compare the inferred values of cosmological parameters from the simulated data vectors to the fiducial values. Following DESY1MG, we adopt $1\sigma$ as the threshold for a bias. If we observe a bias shifted more than $1\sigma$ from the fiducial quantity, the corresponding effect of systematics or modeling assumption needs to be corrected or adequately accounted for in our analysis pipeline. 

The modeling assumptions that we consider are summarized below:

\begin{enumerate}[leftmargin=*, label=(\roman*)]
  \itemsep1em
  \item Limber approximation and RSD: 
the Limber approximation \citep{limber1, limber2} enables us to compute the angular two-point statistics efficiently by simplifying the Bessel function integrals. 
However, this approximation may not be sufficient if the constraining power from surveys can no longer tolerate the errors induced by the approximation. We simulate a data vector using the exact $w(\theta)$ calculation and include the contribution from redshift space distortions \citep{Padmanabhan2007}. 
  
 \item Magnification: the observed number density of foreground galaxies can be altered due to the matter between the foreground galaxies and the observer, leading to a systematic bias to the two-point statistics. For the foreground redshift $z > 0.45$, the effect can bias the inferred dark energy equation of state by more than $5\%$ \citep{Ziour2008}. Accounting for this effect is critical for the higher redshift samples, such as DMASS. To estimate the impact of the magnification bias, we simulate a data vector by injecting the contribution of magnification to $\gamma_t(\theta)$ and $w(\theta)$. 
We assume that the change in the observed galaxy over-density for lenses produced by magnification is proportional to the convergence field of lenses. This can be expressed as $\delta_g^{\rm obs} = \delta_g^{\rm int} + 2(\alpha - 1)  \kappa$, where $\delta_g^{\rm int}$ is the galaxy over-density that would have been observed in the absence of magnification, $\alpha$ is the logarithmic slope of the luminosity function of lenses at its faint end, and $\kappa$ is the convergence of the lenses \citep{Bartelmann2001}. We adopt the value of $\alpha = 2.62$ computed using the CMASS galaxy mocks in \cite{Joachimi2020}. 

\item Intrinsic alignments: our fiducial pipeline accounts for the tidal alignment of massive elliptical galaxies on large scales. However, the contribution from ``tidal torquing''  induced by spiral or less luminous elliptical galaxies is non-negligible on smaller scales. We simulate a data vector using the Tidal Alignment and Tidal Torquing model (TATT) described in \cite{IATATT}. We set the TATT amplitudes to $A_1=0, ~ A_2=2$ with no $z$ dependence as done in \cite{DESCollaboration2017} and DESY1MG. This choice of TATT parameters corresponds to testing the impact of a pure tidal torquing model when analyzed with a tidal alignment model.
\end{enumerate}
Since we cut out the majority of the nonlinear scales as described in Section \ref{sec:data.scalecut}, we do not consider systematics that particularly affects small scales such as baryonic feedback effects. We do not test for the nonlinear galaxy bias either as the linearity of the galaxy bias of DMASS on galaxy clustering and galaxy-galaxy lensing has been verified in \cite{DMASS} and \cite{DMASS-GGL} down to the scales of $4\hMpc$.
These tests have been performed by jointly fitting the two-point statistics with other external probes. 
DESY1MG tested the aforementioned systematics using their 3x2pt simulated data vector and external data sets. For the case of the 3x2pt simulated data vector only, their posteriors of modified gravity were skewed from true values due to the interplay of weak constraining power with a relatively flat likelihood profile and prior volume effect. As we use only one lens bin compared to  DES Y1 five bins, we should expect a worse bias in the posteriors with weaker constraining power. Therefore, we perform the systematics tests by jointly fitting the two-point statistics with the simulated BOSS CMASS likelihood. 
Since the simulated BOSS CMASS likelihood is not sensitive to any of the systematics correlated with the two-point functions, bias from the fiducial values is solely originating from the two-point functions of DMASS. 
We also perform the systematics tests including the actual {\it Planck} data set as shifts by systematics may appear with higher constraining power. 

The results of the robustness tests are shown in Figure \ref{fig:sys-mg}, for the two modified gravity parameters ($\mu_0$, $\Sigma_0$) and the parameters that are most sensitive to in this work: ${\Omega_m, S_8, b}$. 
The top panel shows the constraints obtained from the baseline case, and the bottom panel shows the baseline+{\it Planck} case.
The different rows correspond to the different systematics described above. The topmost error bars in each panel are our fiducial case. For all cases, our posteriors for these cosmological parameters remain well within $1\sigma$ of the fiducial constraints.

Throughout the robustness tests, we use an importance sampling pipeline \citep{WeaverdyckPrep} to rapidly compute posterior means for each systematics. 
For a given sample from a proposal distribution, the importance sampling pipeline draws the expected values of parameters under a target distribution by weighting points with density ratios of the proposal distribution and target distribution, called importance weights. 
We adopt the effective number of samples as a diagnostic tool to evaluate the accuracy of importance sampling, which is given as $N^{\rm IS}_{\rm eff} =  1 / \sum^{N}_{i=1} w_i^2$, where $N$ is the total sample size and $w_i$ represents  normalized importance weights. 
If the proposal distribution is equal to the target distribution, the quantity becomes $N^{\rm IS}_{\rm eff} = N$. 
We determine $N^{\rm IS}_{\rm eff}/N > 0.7$ as the threshold of reliability and evaluate the quantity for each systematics chain obtained by the importance sampler. We find that all of the systematics chains pass the threshold except for the case of intrinsic alignments. Therefore, we use \multinest\ for the intrinsic alignments and utilize the importance sampler for the rest of the systematics.

\begin{figure*}
\centering
\includegraphics[align=t,width=0.34\textwidth]{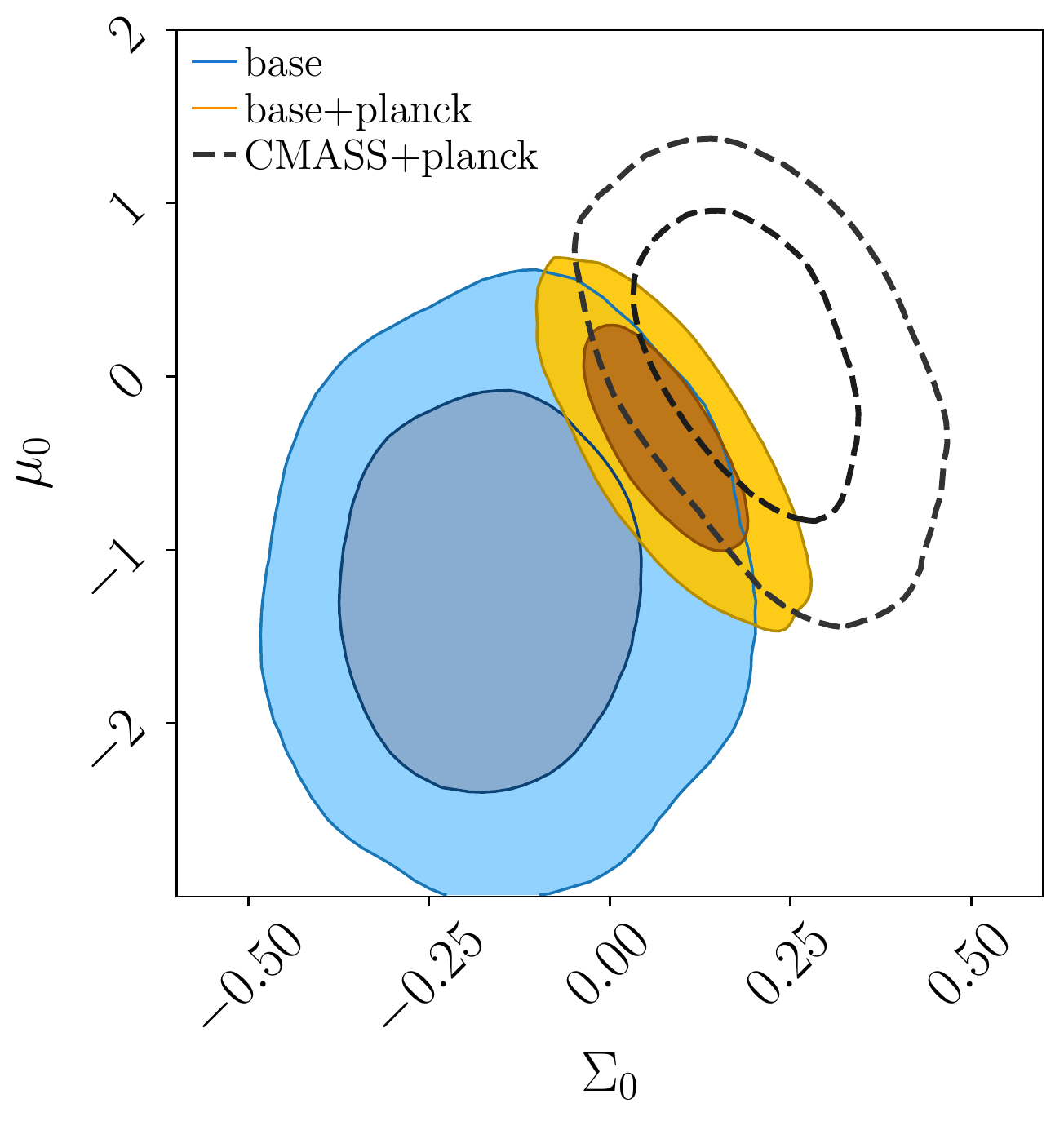}
\includegraphics[align=t,width=0.345\textwidth]{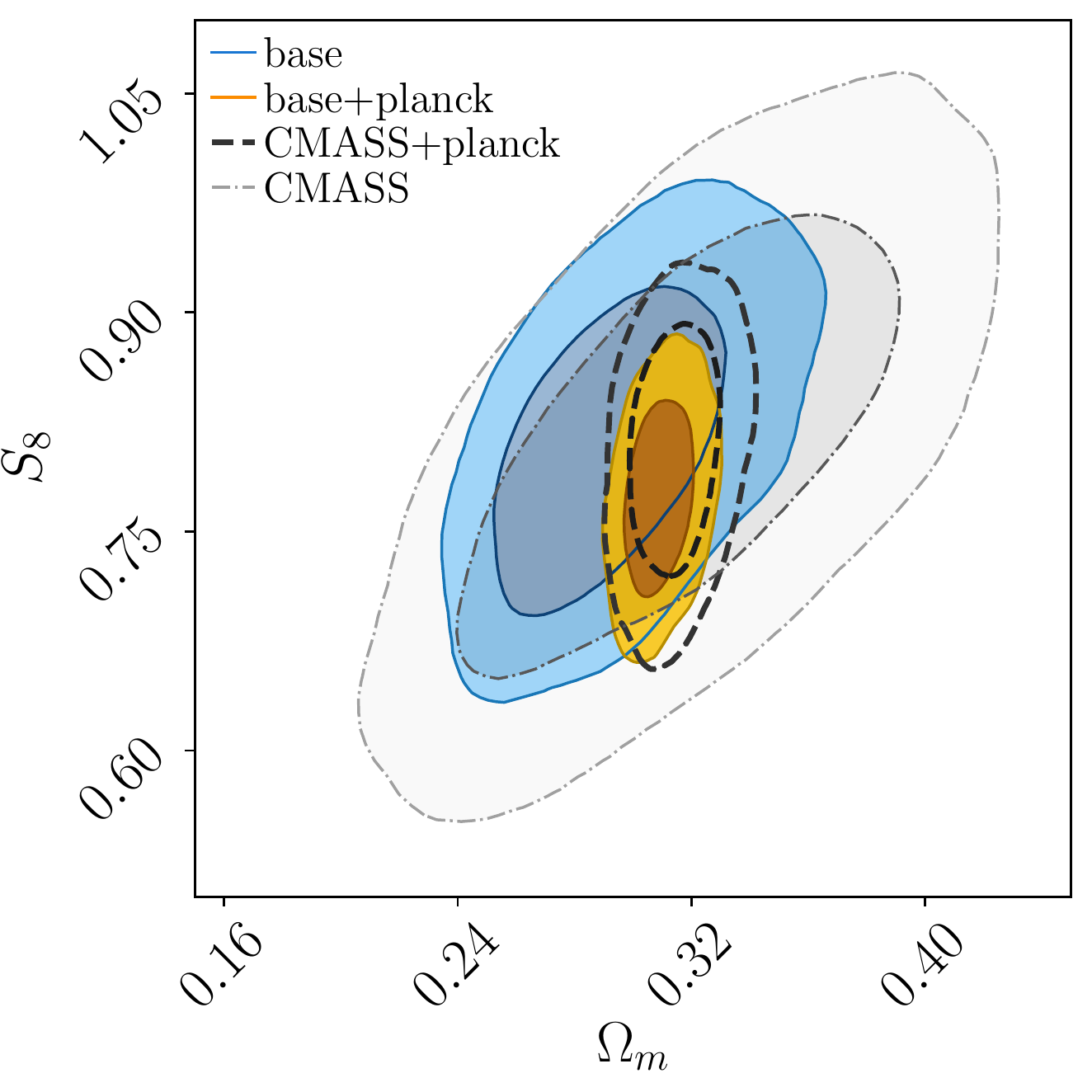}
\caption{ 
Constraints in the $\mu_0-\Sigma_0$ plane (left), and $\Omega_m-S_8$ plane (right) obtained from various combinations of the probes. The blue contours show the constraints obtained from the baseline case (DMASS 3x2pt + CMASS RSD/BAO). The orange contours are obtained by adding the {\it Planck} data to the baseline case. The black-dashed contours are obtained from the external data only. The grey contour in the right panel shows the constraints of $\Omega_m$ and $S_8$ obtained from CMASS only. Adding the DMASS 3x2pt data improves the constrains from the external data sets by $28\%$ for $\mu_0$, $30\%$ for $\Sigma_0$, and $24\%$ for $S_8$. 
 }
\label{fig:mg_result}
\end{figure*}

\section{Blinding}
\label{sec:blinding}

We blinded our results in the parameter-level to avoid confirmation bias. 
The measurements of $\gamma_t$ from  \cite{DMASS-GGL} were blinded in the real data analysis. No comparison to theoretical predictions at the two-point level was made in the blinding stage. The cosmological parameter constraints were scaled and shifted by a random number when plotting.  After ensuring that there were no major systematics that can bias the cosmological constraints through a series of robustness tests listed in Section \ref{sec:robustness}, we unblinded both the data vector of $\gamma_t$ and cosmological constraints. 
As described in the next section, after unblinding we performed an additional test to estimate the impact of the high signal of $\gamma_t$ measured with the first source bin that was detected after unblinding. 
No change was made in either the analysis method or pipeline after unblinding.

\section{Results}
\label{sec:result}

\begin{figure}
\centering
\includegraphics[width=0.23\textwidth]{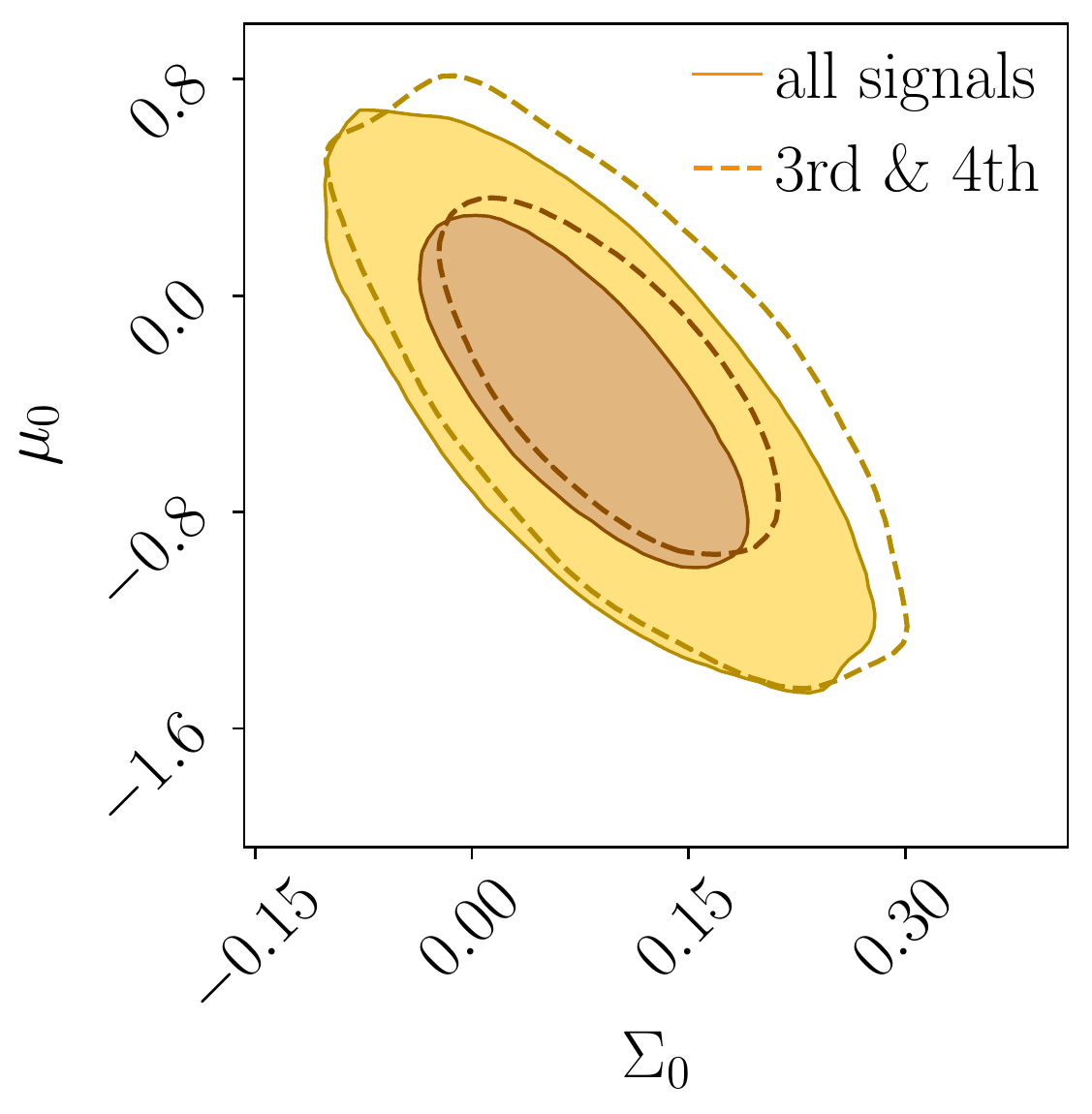}
\includegraphics[width=0.23\textwidth]{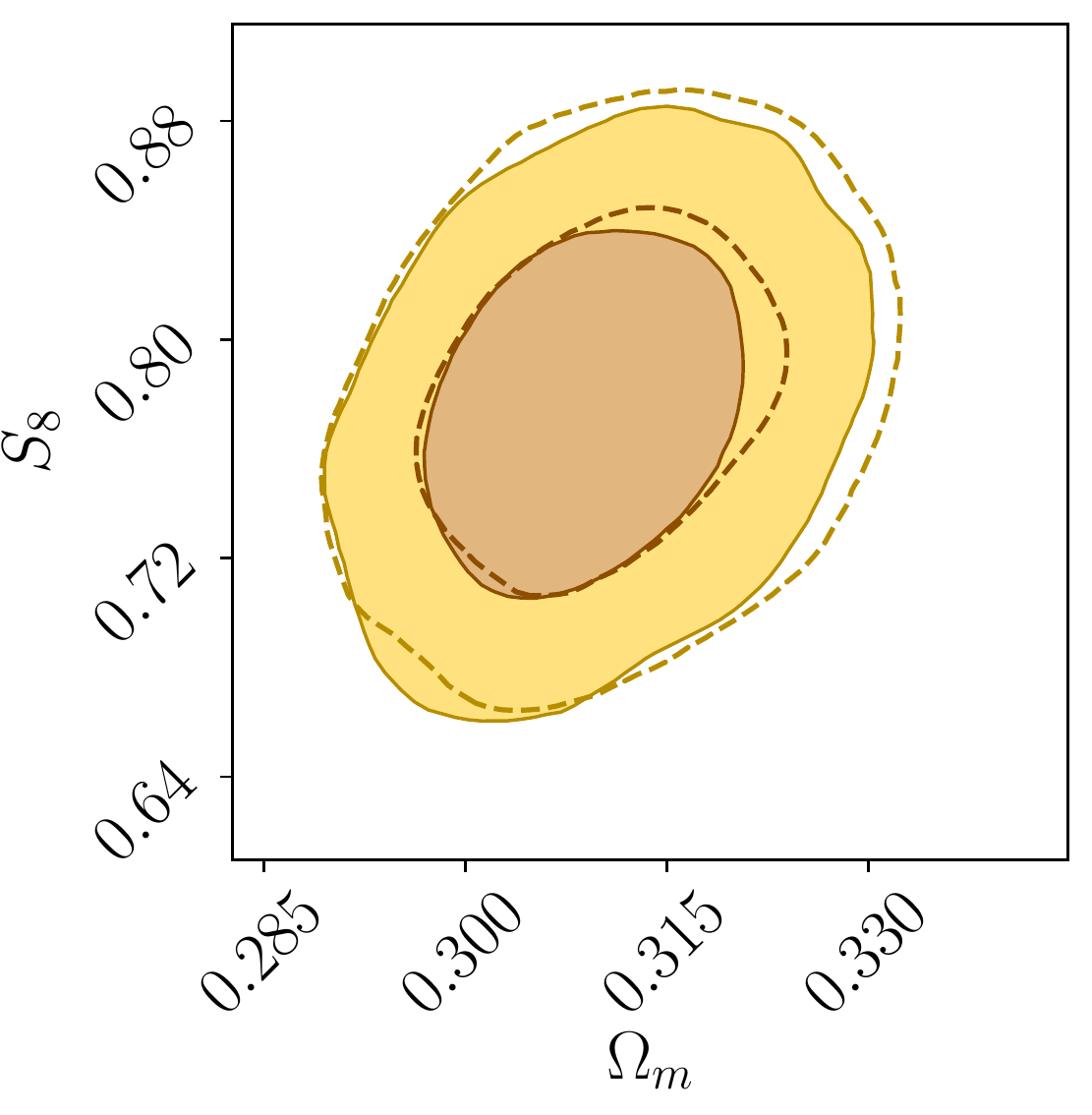}
\caption{ 
Constraints obtained with all four tomographic cross-correlations of $\gamma_t$ (solid) and only with the third and fourth cross-correlations of $\gamma_t$ (dashed) from the two highest source bins. All contours are the case of baseline+{\it Planck}. There is no significant bias or degradation found in the constraints. 
 }
\label{fig:mg_result_src34}
\end{figure}

\begin{figure*}
\centering
\includegraphics[width=0.6\textwidth]{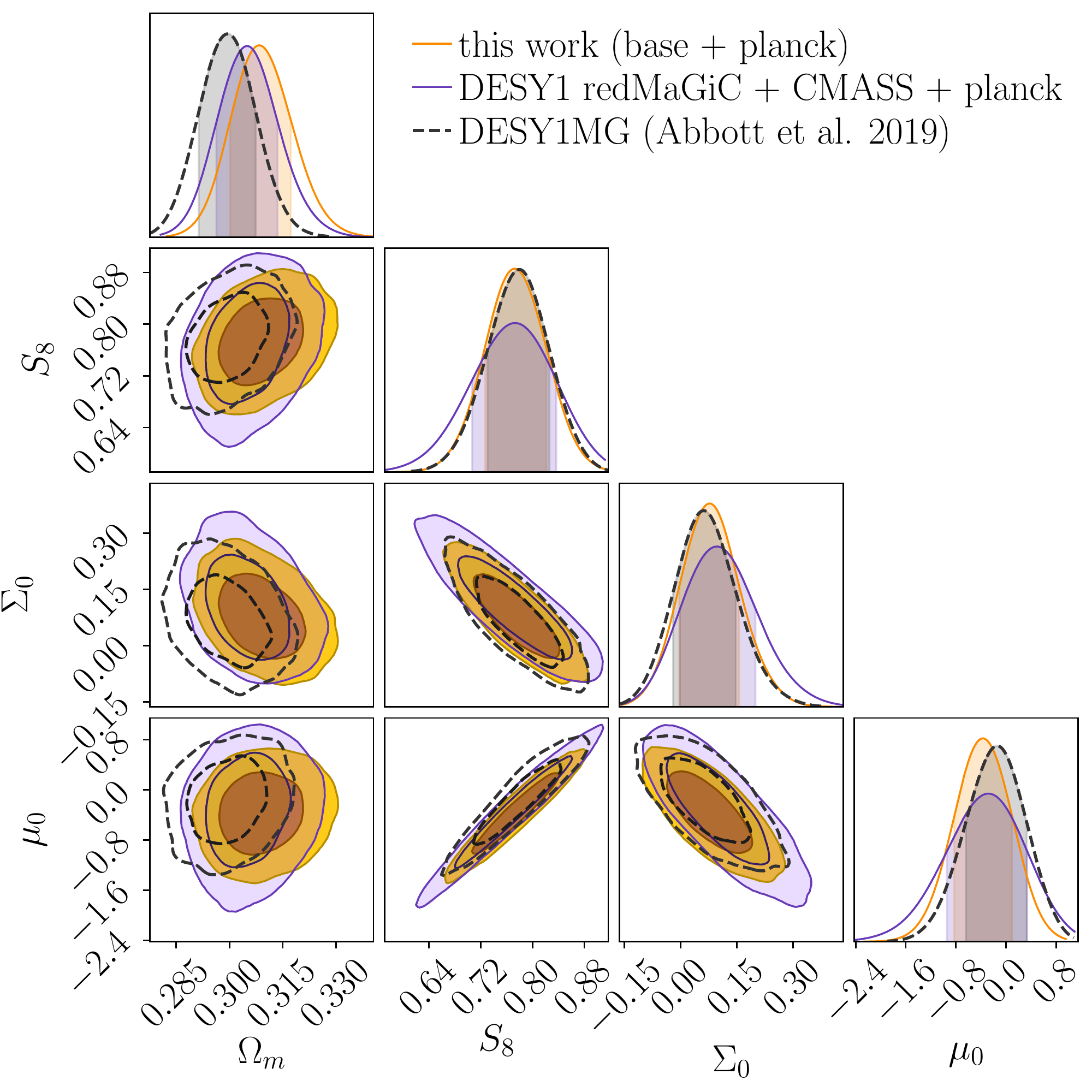}
\caption{ 
Constraints from this work (baseline+{\it Planck}; orange), DESY1 \redmagic\ 3x2pt + CMASS RSD/BAO + {\it Planck} (purple), and the published constraints of the DES Y1 MG analysis (black-dashed) \protect\citep{DES-EXT}. Compared to the case with \redmagic\ (purple), using DMASS (orange) improves the MG constraints by $29\%$ for $\mu_0$ and $21\%$ for $\Sigma_0$ with the same external data sets. The DMASS case is even comparable to that of DESY1MG (black-dashed) despite the lack of the external data sets. The shift of $\Omega_m$ in DESY1MG is caused by the SNe Ia data favoring low $\Omega_m$. Overall, the constraints in the three cases are consistent. 
}
\label{fig:comparison_y1}
\end{figure*}

\subsection{Constraints on modified gravity}

In this section, we show the resulting MG constraints obtained from the various combinations of the data sets listed in Section \ref{sec:data}. For all cases reported, we include the CMASS likelihood to DES 3x2pt as default and do not report the result from DMASS alone because 1) DMASS is designed to harness its maximum power when its measurements are combined with the measurements of CMASS, and 2) DMASS only, without the help of external data sets, suffers from a projection bias in the posteriors of modified gravity as stated in Section \ref{sec:robustness}. 

Figure \ref{fig:mg_result} shows the resulting constraints in the $\mu_0$--$\Sigma_0$ plane (left), and $\Omega_m$--$S_8$ plane (right). Table \ref{tab:results} presents a more detailed summary on our findings, including the constraint of galaxy bias. In Figure \ref{fig:mg_result}, the blue contours show the constraints obtained from the baseline case (DMASS 3x2pt + CMASS RSD/BAO). The resulting values show a mild deviation from GR with $\mu_0 = -1.23^{+0.81}_{-0.83}$ and $\Sigma_0 = -0.17^{+0.16}_{-0.15}$. As $\Sigma_0$ indicates modifications to GR by relativistic particles, the value of $\Sigma_0$ here is driven by weak lensing in DES. 
The orange contours are obtained by adding the CMB data from {\it Planck} to the baseline case. As shown in the figure, adding {\it Planck} recovers GR,  
yielding the values of $\mu_0 = -0.37^{+0.47}_{-0.45}$ and $\Sigma_0 = 0.078^{+0.078}_{-0.082}$. 
Along with these results, we also show the results from the external data only (CMASS RSD/BAO + {\it Planck}; black-dashed) and CMASS RSD/BAO only (grey-dashed) to infer contributions from the DMASS 3x2pt data indirectly. We find that the DMASS 3x2pt data improves the constraints from the external data sets by $28\%$ for $\mu_0$, $30\%$ for $\Sigma_0$, and $24\%$ for $S_8$. For the case of CMASS RSD/BAO alone, $\Sigma_0$ is not constrained by RSD and $\mu_0$ is barely constrained within the imposed prior range. Therefore, we only look into the constraints of $S_8$ and $\Omega_m$. 
Comparing with the baseline, we find that adding DMASS 3x2pt to CMASS RSD/BAO effectively reduces errors on constraints even without {\it Planck}, improving $S_8$ by $45 \%$ and $\Omega_m$ by $32 \%$.   
These notable improvements by adding DMASS 3x2pt are mainly driven by the galaxy bias parameter shared and constrained by DMASS and CMASS simultaneously. The galaxy bias from the baseline+{\it Planck} case is improved by $\sim 80\%$ compared to that of the external data alone, which leads to a significant improvement in other parameters that are degenerate with galaxy bias.

Finally, we test the robustness of our results against unmodeled systematics. As shown in Figure \ref{fig:datavector1}, the tangential shear signal measured with DMASS and the first source bin of DES Y1 is higher than predicted in theory.
\cite{DMASS-GGL} suggests the local peak in galaxy bias at the low-redshift end of the original BOSS CMASS sample as a potential reason for this. If the DMASS algorithm faithfully replicated the properties of CMASS, this peak would have been  replicated in DMASS and even amplified at the low-redshifts edge where the algorithm poorly works. As the first source bin lies largely in front of the lens bin, it is possible that the tangential shear signal measured with the first source bin only captures correlations with the DMASS galaxies at low redshifts where galaxy bias peaks.
%
\cite{DMASS-GGL} tested the impact of the unmodeled local peak on the galaxy bias constraint inferred from the tangential shear measurement. In fixed $\Lambda$CDM cosmology, they found that the resulting constraint is not biased by the local peak above the scale of $4 \hMpc$. It is less likely that the local peak will affect our result as we utilize the scales larger than $4 \hMpc$, and include the angular galaxy clustering and shear measurements that are less sensitive to an irregularity in galaxy bias at the redshift tails. However, it is worth testing its impact in the frame of modified gravity as well to rule out the possibility of this unmodeled systematics biasing the result. For this, we repeat the analysis without the tangential shear signals from the first and second source bins. 
%
%
The signal from the second source bin is excluded to completely remove any potential effect related to galaxy bias from the low-redshift end of DMASS.  
The results are presented in Figure \ref{fig:mg_result_src34}. Each panel shows the contours of the MG parameters (left) and $\Omega_m$--$S_8$ (right) with all tangential shear signals (solid lines) and without the signals from the first two source bins (dashed lines). Both panels are for the case of baseline+{\it Planck}. 
As shown in this figure, we do not find significant bias or degradation in the constraints.  

\subsection{Comparison with the DES Y1 MG analysis}

Earlier, the DES Collaboration reported the constraints on $\mu_0$ and $\Sigma_0$ obtained from the DES Y1 data in DESY1MG. As this work follows their analysis methods with the same shear catalog, comparing the results of this work with DESY1MG enables us to validate the capability of using DMASS as gravitational lenses to combine the spectroscopic and imaging surveys efficiently. This section briefly describes the lens sample and external data sets used in DESY1MG and compares their MG constraints with ours. As we do not report the result for the DES data alone, we only compare the results measured with the external data sets.  
 
DESY1MG utilized the 3x2pt data vectors measured with the \redmagic\ lenses \citep*{Rozo2016RedMaGiC:Data,ELVINPOOLE} and the DES Y1 \metacal\ sources \citep*{METACAL1,METACAL2,Zuntz2018}. The \redmagic\ sample consists of $660,000$ galaxies over an area of $1321\deg^2$. The sample was divided into five redshift bins, using three different cuts on intrinsic luminosity: $0.15 < z < 0.3$, $0.3 < z < 0.45$, $0.45 < z < 0.6$, $0.6 < z < 0.75$, and $0.75 < z < 0.9$. The first three bins were selected using a luminosity cut of $L > 0.5 L_*$ with a spatial density $\bar{n} = 10^{-3} (\hMpc)^{-3}$, the other two bins were selected using $L > L_*$ with $\bar{n}=4 \times 10^{-4} (\hMpc)^{-3}$, and $L > 1.5 L_*$ with $\bar{n}= 10^{-4} (\hMpc)^{-3}$, respectively. 
The redshift coverage of DMASS is comparable to the third and fourth bins of \redmagic\ combined, but the number density of DMASS ($\bar{n} = 3 \times 10^{-4}$) is much lower than \redmagic.
The external data sets selected in DESY1MG were the CMB measurements from {\it Planck} 2015 \citep{Planck2015CosmologicalParameters}, the RSD and BAO measurements from BOSS DR12 \citep{Alam2017}, and the additional BAO measurements from Six-degree Field Galaxy Survey \citep[6dF;][]{6df}, SDSS Main Galaxy Samples \citep[SDSS MGS;][]{SDSS-MGS}, and lastly, Type Ia supernova measurements from Pantheon \citep{Scolnic2017}. 
The resulting MG constraints from these data sets are 
$\mu_0=-0.11^{+0.42}_{-0.46}$ and $\Sigma_0=0.06^{+0.08}_{-0.07}$.

For a more direct comparison with DESY1MG, 
we perform an additional analysis using DES Y1 3x2pt (i.e. \redmagic\ lenses) with CMASS RSD/BAO. 
By doing so, we can infer the pure improvement by DMASS while excluding contributions from other external data sets in DESY1MG. The resulting constraints of MG are $\mu_0=-0.29^{+0.63}_{-0.66}$, $\Sigma_0=0.098^{+0.102}_{-0.100}$. 

Figure \ref{fig:comparison_y1} shows the constraints of this work (baseline + {\it Planck}) in orange, \redmagic \ + CMASS RSD/BAO + {\it Planck} in purple, and DESY1MG in black-dashed lines. 
%
%
Compared to the case with \redmagic\ (purple), it is noticeable that using DMASS (orange) improves the MG constraints by $29\%$ for $\mu_0$ and $21\%$ for $\Sigma_0$ with the same external data sets 
despite \redmagic\ consisting of five tomographic bins with a higher number density. 
Our result is also comparable to that of DESY1MG (black-dashed) despite the lack of the external data sets. This can be explained as follows: five tomographic bins of \redmagic\ introduce additional nuisance parameters of galaxy bias for each bin. When combined with the CMASS likelihood, these five galaxy bias parameters  from \redmagic\ and an additional one from CMASS are varied independently, and hence uncertainties from the total six parameters add to the total statistical error budget. 
This especially affects the modified-gravity parameters 
due to the strong degeneracy among the galaxy bias and MG parameters. 
Compared with the other two cases, DESY1MG (black-dashed) slightly shifts towards a lower value of $\Omega_m$. 
The shift is possibly caused by the Pantheon supernova data that favors a relatively lower value of $\Omega_m$ than the rest of the data sets \citep{Scolnic2017}.  
Overall, the constraints obtained with DMASS are consistent with those with \redmagic. We do not find a significant shift in the central values of $\mu_0$ and $\Sigma_0$ among the three cases.

\begin{table*}
\centering
\caption{1D marginalized posteriors obtained from various combinations of data sets. `baseline' denotes the combination of the DMASS 3x2pt data and RSD and BAO measurements from BOSS CMASS. 
}
\label{tab:results}
\resizebox{0.9\textwidth}{!}{%
\renewcommand{\arraystretch}{1.7}
\begin{tabular}{@{}lclcccccccc@{}}
\toprule
 &  & \multicolumn{1}{c}{Data}      & $\mu_0$                 & $\Sigma_0$                & $\Omega_m$                                             & $\sigma_8$                & $S_8$                     & $b$                     &  &  \\ \midrule
 
 &  & baseline           & 
$-1.23^{+0.81}_{-0.83}$ & $-0.17^{+0.16}_{-0.15}$ & $0.289^{+0.031}_{-0.028}$ & $0.821^{+0.067}_{-0.068}$ & $0.809^{+0.080}_{-0.079}$ & $1.83^{+0.20}_{-0.17}$    &  &  \\
 
 &  & baseline + {\it Planck}  & 
$-0.37^{+0.47}_{-0.45}$ & $0.078^{+0.078}_{-0.082}$ & $\left( 308.2^{+9.0}_{-8.0} \right) \times 10^{-3}$ & $0.762\pm 0.045$ & $0.772^{+0.050}_{-0.046}$ & $2.134^{+0.079}_{-0.078}$  &  &  \\
 
 &  & DESY1 \redmagic\ + CMASS + {\it Planck} & 
$-0.29^{+0.63}_{-0.66}$ & $0.098^{+0.102}_{-0.100}$ & $\left( 305.0\pm 8.6 \right) \times 10^{-3}$ & $0.767^{+0.061}_{-0.065}$ & $0.773^{+0.064}_{-0.067}$ & $-$                       &  &  \\
 
 &  & CMASS + {\it Planck}                & 
$0.15^{+0.61}_{-0.67}$ & $0.20^{+0.12}_{-0.11}$ & $0.314^{+0.012}_{-0.011}$ & $0.794^{+0.057}_{-0.062}$ & $0.814^{+0.059}_{-0.068}$ & $2.00^{+0.28}_{-0.26}$     &  & \\

 &  & CMASS                & 
$-$ & $-$ & $0.319^{+0.052}_{-0.055}$ & $0.792^{+0.087}_{-0.084}$ & $0.819^{+0.114}_{-0.119}$ & $1.980^{+0.348}_{-0.313}$     &  &

 \\ \bottomrule
\end{tabular}%
}
\end{table*}

\section{Conclusion}
\label{sec:conclusion}

In this paper, we measured the modified gravity parameters $\Sigma_0$ and $\mu_0$ using the DMASS sample to demonstrate its feasibility as an equivalent to BOSS CMASS in the DES Y1 footprint for a joint analysis of BOSS and DES. We utilized the angular clustering of the DMASS galaxies, cosmic shear of the DES \metacal\ sources, and cross-correlation of the two as data vectors. Then we derived the constraints on MG by jointly fitting the data vectors with the RSD and BAO measurements from the BOSS CMASS sample and CMB measurements from {\it Planck}. 
As detailed in Section \ref{sec:analysis}, we thoroughly followed the analysis methods of DESY1MG. We employed the same modeling for the two-point statistics and potential systematics, such as the shear and photo-z biases. 
A series of tests were performed as done in DESY1MG to make sure our results are robust to the modeling assumptions that are not included in the pipeline. 

The resulting constraints are consistent with GR. Adding DMASS 3x2pt significantly improves the existing constraints from CMASS RSD/BAO + {\it Planck} by $28\%$ for $\mu_0$ and $30\%$ for $\Sigma_0$. Not only that, we found that DMASS yields constraints improved by $29\%$ for $\mu_0$ and $21\%$ for $\Sigma_0$, compared to the constraints obtained with \redmagic\ when the same external data sets are utilized. 
The results are also comparable to the results of DESY1MG performed with additional BAO \& RSD and supernova measurements. 
This is mainly because of a tightly constrained galaxy bias parameter shared in the CMASS and DMASS likelihoods, which breaks the degeneracy between galaxy bias and other parameters. Compared to that, the cases with \redmagic\ treats the galaxy bias parameter from RSD as an additional one, which varies independently from those of \redmagic. The uncertainty induced by the independent use of galaxy bias erases gains from additional BAO \& RSD and supernova measurements.

This approach to optimally combine the DES and BOSS surveys with DMASS can be easily extended to other types of spectroscopic samples and image-based surveys. 
In just a few years, data from the Stage IV surveys will become available. The survey footprint of LSST\footnote{Rubin Observatory's Legacy Survey of Space and Time} \citep{LSST} will occupy the entire southern sky with a small overlapping area with spectroscopic surveys such as DESI\footnote{Dark Energy Spectroscopic Instrument} \citep{DESI2016} in the northern sky. 
The DESI survey produces the Bright Galaxy sample (BGS) at $z < 0.4$ and the Emission Line Galaxy sample (ELG) at $z > 0.6$. The equivalent samples to those DESI targets in the LSST area will be ideal lenses to extract the lensing signals in LSST that can be efficiently combined with the galaxy clustering of the DESI galaxy samples as done with DMASS and CMASS. Combining the two surveys in this manner will embrace almost the entire sky and yield a wealth of information to test gravity.

\section*{Acknowledgements}

We thank D. H. Weinberg for useful conversations in the course of preparing this work. 

AC acknowledges support from NASA grant 15-WFIRST15-0008. During the preparation of this paper, C.H.\ was supported by the Simons Foundation, NASA, and the US Department of Energy.

The figures in this work are produced with plotting routines from matplotlib \citep{matplotlib} and ChainConsumer \citep{ChainConsumer}.
Some of the results in this paper have been derived using the healpy and HEALPix package \citep{HEALPix,healpy}.

Funding for the DES Projects has been provided by the U.S. Department of Energy, the U.S. National Science Foundation, the Ministry of Science and Education of Spain, 
the Science and Technology Facilities Council of the United Kingdom, the Higher Education Funding Council for England, the National Center for Supercomputing 
Applications at the University of Illinois at Urbana-Champaign, the Kavli Institute of Cosmological Physics at the University of Chicago, 
the Center for Cosmology and Astro-Particle Physics at the Ohio State University,
the Mitchell Institute for Fundamental Physics and Astronomy at Texas A\&M University, Financiadora de Estudos e Projetos, 
Funda{\c c}{\~a}o Carlos Chagas Filho de Amparo {\`a} Pesquisa do Estado do Rio de Janeiro, Conselho Nacional de Desenvolvimento Cient{\'i}fico e Tecnol{\'o}gico and 
the Minist{\'e}rio da Ci{\^e}ncia, Tecnologia e Inova{\c c}{\~a}o, the Deutsche Forschungsgemeinschaft and the Collaborating Institutions in the Dark Energy Survey. 

The Collaborating Institutions are Argonne National Laboratory, the University of California at Santa Cruz, the University of Cambridge, Centro de Investigaciones Energ{\'e}ticas, 
Medioambientales y Tecnol{\'o}gicas-Madrid, the University of Chicago, University College London, the DES-Brazil Consortium, the University of Edinburgh, 
the Eidgen{\"o}ssische Technische Hochschule (ETH) Z{\"u}rich, 
Fermi National Accelerator Laboratory, the University of Illinois at Urbana-Champaign, the Institut de Ci{\`e}ncies de l'Espai (IEEC/CSIC), 
the Institut de F{\'i}sica d'Altes Energies, Lawrence Berkeley National Laboratory, the Ludwig-Maximilians Universit{\"a}t M{\"u}nchen and the associated Excellence Cluster Universe, 
the University of Michigan, NFS's NOIRLab, the University of Nottingham, The Ohio State University, the University of Pennsylvania, the University of Portsmouth, 
SLAC National Accelerator Laboratory, Stanford University, the University of Sussex, Texas A\&M University, and the OzDES Membership Consortium.

Based in part on observations at Cerro Tololo Inter-American Observatory at NSF's NOIRLab (NOIRLab Prop. ID 2012B-0001; PI: J. Frieman), which is managed by the Association of Universities for Research in Astronomy (AURA) under a cooperative agreement with the National Science Foundation.

The DES data management system is supported by the National Science Foundation under Grant Numbers AST-1138766 and AST-1536171.
The DES participants from Spanish institutions are partially supported by MICINN under grants ESP2017-89838, PGC2018-094773, PGC2018-102021, SEV-2016-0588, SEV-2016-0597, and MDM-2015-0509, some of which include ERDF funds from the European Union. IFAE is partially funded by the CERCA program of the Generalitat de Catalunya.
Research leading to these results has received funding from the European Research
Council under the European Union's Seventh Framework Program (FP7/2007-2013) including ERC grant agreements 240672, 291329, and 306478.
We  acknowledge support from the Brazilian Instituto Nacional de Ci\^encia
e Tecnologia (INCT) do e-Universo (CNPq grant 465376/2014-2).

This manuscript has been authored by Fermi Research Alliance, LLC under Contract No. DE-AC02-07CH11359 with the U.S. Department of Energy, Office of Science, Office of High Energy Physics.

Funding for SDSS-III has been provided by the Alfred P. Sloan Foundation, the Participating Institutions, the National Science Foundation, and the U.S. Department of Energy Office of Science. The SDSS-III web site is \url{http://www.sdss3.org/}.

SDSS-III is managed by the Astrophysical Research Consortium for the Participating Institutions of the SDSS-III Collaboration including the University of Arizona, the Brazilian Participation Group, Brookhaven National Laboratory, Carnegie Mellon University, University of Florida, the French Participation Group, the German Participation Group, Harvard University, the Instituto de Astrofisica de Canarias, the Michigan State/Notre Dame/JINA Participation Group, Johns Hopkins University, Lawrence Berkeley National Laboratory, Max Planck Institute for Astrophysics, Max Planck Institute for Extraterrestrial Physics, New Mexico State University, New York University, Ohio State University, Pennsylvania State University, University of Portsmouth, Princeton University, the Spanish Participation Group, University of Tokyo, University of Utah, Vanderbilt University, University of Virginia, University of Washington, and Yale University.

This work used resources at the Owens Cluster at the Ohio Supercomputer Center  \citep{OSC} and the Duke Compute Cluster (DCC) at Duke University.

\section*{Data Availability}

The DMASS galaxy catalog used in this work and some of the ancillary data products shown in the Figures are available through the DES data release page (\url{https://des.ncsa.illinois.edu/releases/other/y1-dmass}). Readers interested in comparing to these results are encouraged to check out the release page or contact the corresponding author for additional information.

\bibliography{dmass_mg_reference,dmass_ggl_reference}

\begin{thebibliography}{}
\makeatletter
\relax
\def\mn@urlcharsother{\let\do\@makeother \do\$\do\&\do\#\do\^\do\_\do\%\do\~}
\def\mn@doi{\begingroup\mn@urlcharsother \@ifnextchar [ {\mn@doi@}
  {\mn@doi@[]}}
\def\mn@doi@[#1]#2{\def\@tempa{#1}\ifx\@tempa\@empty \href
  {http://dx.doi.org/#2} {doi:#2}\else \href {http://dx.doi.org/#2} {#1}\fi
  \endgroup}
\def\mn@eprint#1#2{\mn@eprint@#1:#2::\@nil}
\def\mn@eprint@arXiv#1{\href {http://arxiv.org/abs/#1} {{\tt arXiv:#1}}}
\def\mn@eprint@dblp#1{\href {http://dblp.uni-trier.de/rec/bibtex/#1.xml}
  {dblp:#1}}
\def\mn@eprint@#1:#2:#3:#4\@nil{\def\@tempa {#1}\def\@tempb {#2}\def\@tempc
  {#3}\ifx \@tempc \@empty \let \@tempc \@tempb \let \@tempb \@tempa \fi \ifx
  \@tempb \@empty \def\@tempb {arXiv}\fi \@ifundefined
  {mn@eprint@\@tempb}{\@tempb:\@tempc}{\expandafter \expandafter \csname
  mn@eprint@\@tempb\endcsname \expandafter{\@tempc}}}

\bibitem[\protect\citeauthoryear{Abbott et~al.}{Abbott
  et~al.}{2005}]{DESCollaboration2006}
Abbott T.,  et~al., 2005, arXiv e-prints, \href
  {https://ui.adsabs.harvard.edu/abs/2005astro.ph.10346T} {pp
  astro--ph/0510346}

\bibitem[\protect\citeauthoryear{{Abbott} \& {Abdalla} et~al.,}{{Abbott}
  et~al.}{2016}]{DESOverview}
{Abbott} T.,  et~al. 2016, \mn@doi [\mnras] {10.1093/mnras/stw641}, \href
  {https://ui.adsabs.harvard.edu/\#abs/2016MNRAS.460.1270D} {460, 1270}

\bibitem[\protect\citeauthoryear{{Abbott} \& {Abdalla} et~al.,}{{Abbott}
  et~al.}{2018}]{DESCollaboration2017}
{Abbott} T.~M.~C.,  et~al. 2018, \mn@doi [\prd] {10.1103/PhysRevD.98.043526},
  \href {https://ui.adsabs.harvard.edu/\#abs/2018PhRvD..98d3526A} {98, 043526}

\bibitem[\protect\citeauthoryear{{Abbott} \& {Abdalla} et~al.,}{{Abbott}
  et~al.}{2019}]{DES-EXT}
{Abbott} T.~M.~C.,  et~al. 2019, \mn@doi [\prd] {10.1103/PhysRevD.99.123505},
  \href {https://ui.adsabs.harvard.edu/abs/2019PhRvD..99l3505A} {99, 123505}

\bibitem[\protect\citeauthoryear{{Aihara} \& {Arimoto} et~al.,}{{Aihara}
  et~al.}{2018}]{HSCAihara}
{Aihara} H.,  et~al. 2018, \mn@doi [\pasj] {10.1093/pasj/psx066}, \href
  {https://ui.adsabs.harvard.edu/abs/2018PASJ...70S...4A} {70, S4}

\bibitem[\protect\citeauthoryear{{Alam} \& {Albareti} et~al.,}{{Alam}
  et~al.}{2015}]{Alam2015TheSDSS-III}
{Alam} S.,  et~al. 2015, \mn@doi [The Astrophysical Journal Supplement Series]
  {10.1088/0067-0049/219/1/12}, \href
  {https://ui.adsabs.harvard.edu/\#abs/2015ApJS..219...12A} {219, 12}

\bibitem[\protect\citeauthoryear{{Alam} \& {Miyatake} et~al.,}{{Alam}
  et~al.}{2017a}]{Alam2017TestingCMASS}
{Alam} S.,  et~al. 2017a, \mn@doi [\mnras] {10.1093/mnras/stw3056}, \href
  {https://ui.adsabs.harvard.edu/\#abs/2017MNRAS.465.4853A} {465, 4853}

\bibitem[\protect\citeauthoryear{{Alam} \& {Ata} et~al.,}{{Alam}
  et~al.}{2017b}]{Alam2017}
{Alam} S.,  et~al. 2017b, \mn@doi [\mnras] {10.1093/mnras/stx721}, \href
  {https://ui.adsabs.harvard.edu/abs/2017MNRAS.470.2617A} {470, 2617}

\bibitem[\protect\citeauthoryear{{Amon} \& {Blake} et~al.,}{{Amon}
  et~al.}{2018}]{Amon2018MG}
{Amon} A.,  et~al. 2018, \mn@doi [\mnras] {10.1093/mnras/sty1624}, \href
  {https://ui.adsabs.harvard.edu/\#abs/2018MNRAS.479.3422A} {479, 3422}

\bibitem[\protect\citeauthoryear{{Bartelmann} \& {Schneider}}{{Bartelmann} \&
  {Schneider}}{2001}]{Bartelmann2001}
{Bartelmann} M.,  {Schneider} P.,  2001, \mn@doi [\physrep]
  {10.1016/S0370-1573(00)00082-X}, \href
  {https://ui.adsabs.harvard.edu/abs/2001PhR...340..291B} {340, 291}

\bibitem[\protect\citeauthoryear{{Bernardeau}}{{Bernardeau}}{1998}]{Bernardeau1998}
{Bernardeau} F.,  1998, \aap, \href
  {https://ui.adsabs.harvard.edu/abs/1998A&A...338..767B} {338, 767}

\bibitem[\protect\citeauthoryear{{Bernstein} \& {Cai}}{{Bernstein} \&
  {Cai}}{2011}]{Bernstein2011}
{Bernstein} G.~M.,  {Cai} Y.-C.,  2011, \mn@doi [\mnras]
  {10.1111/j.1365-2966.2011.19249.x}, \href
  {https://ui.adsabs.harvard.edu/abs/2011MNRAS.416.3009B} {416, 3009}

\bibitem[\protect\citeauthoryear{{Beutler} \& {Blake} et~al.,}{{Beutler}
  et~al.}{2012}]{6df}
{Beutler} F.,  et~al. 2012, \mn@doi [\mnras]
  {10.1111/j.1365-2966.2012.21136.x}, \href
  {https://ui.adsabs.harvard.edu/abs/2012MNRAS.423.3430B} {423, 3430}

\bibitem[\protect\citeauthoryear{{Blake} \& {Joudaki} et~al.,}{{Blake}
  et~al.}{2016}]{Blake2016}
{Blake} C.,  et~al. 2016, \mn@doi [\mnras] {10.1093/mnras/stv2875}, \href
  {https://ui.adsabs.harvard.edu/abs/2016MNRAS.456.2806B} {456, 2806}

\bibitem[\protect\citeauthoryear{{Blazek} \& {MacCrann} et~al.,}{{Blazek}
  et~al.}{2019}]{IATATT}
{Blazek} J.~A.,  et~al. 2019, \mn@doi [\prd] {10.1103/PhysRevD.100.103506},
  \href {https://ui.adsabs.harvard.edu/abs/2019PhRvD.100j3506B} {100, 103506}

\bibitem[\protect\citeauthoryear{{Bolton} \& {Schlegel} et~al.,}{{Bolton}
  et~al.}{2012}]{Bolton2012SpectralSurvey}
{Bolton} A.~S.,  et~al. 2012, \mn@doi [\aj] {10.1088/0004-6256/144/5/144},
  \href {https://ui.adsabs.harvard.edu/\#abs/2012AJ....144..144B} {144, 144}

\bibitem[\protect\citeauthoryear{{Cai} \& {Bernstein}}{{Cai} \&
  {Bernstein}}{2012}]{Cai2012}
{Cai} Y.-C.,  {Bernstein} G.,  2012, \mn@doi [\mnras]
  {10.1111/j.1365-2966.2012.20676.x}, \href
  {https://ui.adsabs.harvard.edu/abs/2012MNRAS.422.1045C} {422, 1045}

\bibitem[\protect\citeauthoryear{{Chuang} \& {Pellejero-Ibanez}
  et~al.,}{{Chuang} et~al.}{2017}]{Chuang2017}
{Chuang} C.-H.,  et~al. 2017, \mn@doi [\mnras] {10.1093/mnras/stx1641}, \href
  {https://ui.adsabs.harvard.edu/\#abs/2017MNRAS.471.2370C} {471, 2370}

\bibitem[\protect\citeauthoryear{{Coe} \& {Ben{\'\i}tez} et~al.,}{{Coe}
  et~al.}{2006}]{BPZ}
{Coe} D.,  et~al. 2006, \mn@doi [\aj] {10.1086/505530}, \href
  {https://ui.adsabs.harvard.edu/abs/2006AJ....132..926C} {132, 926}

\bibitem[\protect\citeauthoryear{{Cooray} \& {Sheth}}{{Cooray} \&
  {Sheth}}{2002}]{Cooray2002}
{Cooray} A.,  {Sheth} R.,  2002, \mn@doi [\physrep]
  {10.1016/S0370-1573(02)00276-4}, \href
  {https://ui.adsabs.harvard.edu/abs/2002PhR...372....1C} {372, 1}

\bibitem[\protect\citeauthoryear{{DESI Collaboration} \& {Aghamousa}
  et~al.,}{{DESI Collaboration}}{2016}]{DESI2016}
{DESI Collaboration} 2016, arXiv e-prints, \href
  {https://ui.adsabs.harvard.edu/abs/2016arXiv161100036D} {p. arXiv:1611.00036}

\bibitem[\protect\citeauthoryear{{Dawson} \& {Schlegel} et~al.,}{{Dawson}
  et~al.}{2013}]{Dawson2013TheSDSS-III}
{Dawson} K.~S.,  et~al. 2013, \mn@doi [\aj] {10.1088/0004-6256/145/1/10}, \href
  {https://ui.adsabs.harvard.edu/\#abs/2013AJ....145...10D} {145, 10}

\bibitem[\protect\citeauthoryear{{Drlica-Wagner} \& {Sevilla-Noarbe}
  et~al.,}{{Drlica-Wagner} et~al.}{2018}]{Y1GOLD}
{Drlica-Wagner} A.,  et~al. 2018, \mn@doi [The Astrophysical Journal Supplement
  Series] {10.3847/1538-4365/aab4f5}, \href
  {https://ui.adsabs.harvard.edu/\#abs/2018ApJS..235...33D} {235, 33}

\bibitem[\protect\citeauthoryear{{Eisenstein} \& {Weinberg}
  et~al.,}{{Eisenstein} et~al.}{2011}]{Eisenstein2011BOSS}
{Eisenstein} D.~J.,  et~al. 2011, \mn@doi [\aj] {10.1088/0004-6256/142/3/72},
  \href {https://ui.adsabs.harvard.edu/\#abs/2011AJ....142...72E} {142, 72}

\bibitem[\protect\citeauthoryear{{Elvin-Poole} \& {Crocce}
  et~al.,}{{Elvin-Poole} et~al.}{2018}]{ELVINPOOLE}
{Elvin-Poole} J.,  et~al. 2018, \mn@doi [\prd] {10.1103/PhysRevD.98.042006},
  \href {https://ui.adsabs.harvard.edu/\#abs/2018PhRvD..98d2006E} {98, 042006}

\bibitem[\protect\citeauthoryear{{Feroz} \& {Hobson}}{{Feroz} \&
  {Hobson}}{2008}]{multinest1}
{Feroz} F.,  {Hobson} M.~P.,  2008, \mn@doi [\mnras]
  {10.1111/j.1365-2966.2007.12353.x}, \href
  {https://ui.adsabs.harvard.edu/abs/2008MNRAS.384..449F} {384, 449}

\bibitem[\protect\citeauthoryear{{Feroz}, {Hobson}  \& {Bridges}}{{Feroz}
  et~al.}{2009}]{multinest2}
{Feroz} F.,  {Hobson} M.~P.,   {Bridges} M.,  2009, \mn@doi [\mnras]
  {10.1111/j.1365-2966.2009.14548.x}, \href
  {https://ui.adsabs.harvard.edu/abs/2009MNRAS.398.1601F} {398, 1601}

\bibitem[\protect\citeauthoryear{{Feroz} \& {Hobson} et~al.,}{{Feroz}
  et~al.}{2019}]{multinest3}
{Feroz} F.,  et~al. 2019, \mn@doi [The Open Journal of Astrophysics]
  {10.21105/astro.1306.2144}, \href
  {https://ui.adsabs.harvard.edu/abs/2019OJAp....2E..10F} {2, 10}

\bibitem[\protect\citeauthoryear{{Ferreira} \& {Skordis}}{{Ferreira} \&
  {Skordis}}{2010}]{Ferreira2010}
{Ferreira} P.~G.,  {Skordis} C.,  2010, \mn@doi [\prd]
  {10.1103/PhysRevD.81.104020}, \href
  {https://ui.adsabs.harvard.edu/abs/2010PhRvD..81j4020F} {81, 104020}

\bibitem[\protect\citeauthoryear{{Fert{\'e}} \& {Kirk} et~al.,}{{Fert{\'e}}
  et~al.}{2019}]{Ferte2019}
{Fert{\'e}} A.,  et~al. 2019, \mn@doi [\prd] {10.1103/PhysRevD.99.083512},
  \href {https://ui.adsabs.harvard.edu/abs/2019PhRvD..99h3512F} {99, 083512}

\bibitem[\protect\citeauthoryear{{Flaugher} \& {Diehl} et~al.,}{{Flaugher}
  et~al.}{2015}]{Flaugher2015THECAMERA}
{Flaugher} B.,  et~al. 2015, \mn@doi [\aj] {10.1088/0004-6256/150/5/150}, \href
  {https://ui.adsabs.harvard.edu/\#abs/2015AJ....150..150F} {150, 150}

\bibitem[\protect\citeauthoryear{{Gazta{\~n}aga} \& {Eriksen}
  et~al.,}{{Gazta{\~n}aga} et~al.}{2012}]{Gaztanaga2012}
{Gazta{\~n}aga} E.,  et~al. 2012, \mn@doi [\mnras]
  {10.1111/j.1365-2966.2012.20613.x}, \href
  {https://ui.adsabs.harvard.edu/abs/2012MNRAS.422.2904G} {422, 2904}

\bibitem[\protect\citeauthoryear{{G{\'o}rski} \& {Hivon} et~al.,}{{G{\'o}rski}
  et~al.}{2005}]{HEALPix}
{G{\'o}rski} K.~M.,  et~al. 2005, \mn@doi [\apj] {10.1086/427976}, \href
  {http://adsabs.harvard.edu/abs/2005ApJ...622..759G} {622, 759}

\bibitem[\protect\citeauthoryear{{Heymans} \& {Van Waerbeke} et~al.,}{{Heymans}
  et~al.}{2012}]{CFHTLenS}
{Heymans} C.,  et~al. 2012, \mn@doi [\mnras]
  {10.1111/j.1365-2966.2012.21952.x}, \href
  {https://ui.adsabs.harvard.edu/\#abs/2012MNRAS.427..146H} {427, 146}

\bibitem[\protect\citeauthoryear{{Hinton}}{{Hinton}}{2016}]{ChainConsumer}
{Hinton} S.~R.,  2016, \mn@doi [The Journal of Open Source Software]
  {10.21105/joss.00045}, \href
  {https://ui.adsabs.harvard.edu/abs/2016JOSS....1...45H} {1, 00045}

\bibitem[\protect\citeauthoryear{{Hojjati}, {Pogosian}  \& {Zhao}}{{Hojjati}
  et~al.}{2011}]{mgcamb2}
{Hojjati} A.,  {Pogosian} L.,   {Zhao} G.-B.,  2011, \mn@doi [\jcap]
  {10.1088/1475-7516/2011/08/005}, \href
  {https://ui.adsabs.harvard.edu/abs/2011JCAP...08..005H} {2011, 005}

\bibitem[\protect\citeauthoryear{{Hoyle} \& {Gruen} et~al.,}{{Hoyle}
  et~al.}{2018}]{Hoyle2018}
{Hoyle} B.,  et~al. 2018, \mn@doi [\mnras] {10.1093/mnras/sty957}, \href
  {https://ui.adsabs.harvard.edu/abs/2018MNRAS.478..592H} {478, 592}

\bibitem[\protect\citeauthoryear{{Huff} \& {Mandelbaum}}{{Huff} \&
  {Mandelbaum}}{2017}]{METACAL2}
{Huff} E.,  {Mandelbaum} R.,  2017, arXiv e-prints, \href
  {https://ui.adsabs.harvard.edu/abs/2017arXiv170202600H} {p. arXiv:1702.02600}

\bibitem[\protect\citeauthoryear{{Hunter}}{{Hunter}}{2007}]{matplotlib}
{Hunter} J.~D.,  2007, \mn@doi [Computing in Science and Engineering]
  {10.1109/MCSE.2007.55}, \href
  {https://ui.adsabs.harvard.edu/abs/2007CSE.....9...90H} {9, 90}

\bibitem[\protect\citeauthoryear{{Ishak}}{{Ishak}}{2019}]{Ishak2019}
{Ishak} M.,  2019, \mn@doi [Living Reviews in Relativity]
  {10.1007/s41114-018-0017-4}, \href
  {https://ui.adsabs.harvard.edu/abs/2019LRR....22....1I} {22, 1}

\bibitem[\protect\citeauthoryear{{Ivezi{\'c}} \& {Kahn} et~al.,}{{Ivezi{\'c}}
  et~al.}{2019}]{LSST}
{Ivezi{\'c}} {\v{Z}}.,  et~al. 2019, \mn@doi [\apj] {10.3847/1538-4357/ab042c},
  \href {https://ui.adsabs.harvard.edu/abs/2019ApJ...873..111I} {873, 111}

\bibitem[\protect\citeauthoryear{{Joachimi} \& {Lin} et~al.,}{{Joachimi}
  et~al.}{2020}]{Joachimi2020}
{Joachimi} B.,  et~al. 2020, arXiv e-prints, \href
  {https://ui.adsabs.harvard.edu/abs/2020arXiv200701844J} {p. arXiv:2007.01844}

\bibitem[\protect\citeauthoryear{{\swap{Jong}{de }} \& {Verdoes Kleijn}
  et~al.,}{{\swap{Jong}{de }} et~al.}{2013}]{KIDS}
{\swap{Jong}{de }} J. T.~A.,  et~al. 2013, \mn@doi [Experimental Astronomy]
  {10.1007/s10686-012-9306-1}, \href
  {https://ui.adsabs.harvard.edu/\#abs/2013ExA....35...25D} {35, 25}

\bibitem[\protect\citeauthoryear{{Joudaki} \& {Mead} et~al.,}{{Joudaki}
  et~al.}{2017}]{KIDS450MG}
{Joudaki} S.,  et~al. 2017, \mn@doi [\mnras] {10.1093/mnras/stx998}, \href
  {https://ui.adsabs.harvard.edu/abs/2017MNRAS.471.1259J} {471, 1259}

\bibitem[\protect\citeauthoryear{{Jullo} \& {de la Torre} et~al.,}{{Jullo}
  et~al.}{2019}]{Jullo2019}
{Jullo} E.,  et~al. 2019, \mn@doi [\aap] {10.1051/0004-6361/201834629}, \href
  {https://ui.adsabs.harvard.edu/abs/2019A&A...627A.137J} {627, A137}

\bibitem[\protect\citeauthoryear{{Kaiser}}{{Kaiser}}{1984}]{Kaiser1984}
{Kaiser} N.,  1984, \mn@doi [\apj] {10.1086/184341}, \href
  {https://ui.adsabs.harvard.edu/\#abs/1984ApJ...284L...9K} {284, L9}

\bibitem[\protect\citeauthoryear{{Kaiser}}{{Kaiser}}{1992}]{Kaiser1992}
{Kaiser} N.,  1992, \mn@doi [\apj] {10.1086/171151}, \href
  {https://ui.adsabs.harvard.edu/abs/1992ApJ...388..272K} {388, 272}

\bibitem[\protect\citeauthoryear{{Krause} \& {Eifler}}{{Krause} \&
  {Eifler}}{2017}]{COSMOLIKE}
{Krause} E.,  {Eifler} T.,  2017, \mn@doi [\mnras] {10.1093/mnras/stx1261},
  \href {https://ui.adsabs.harvard.edu/abs/2017MNRAS.470.2100K} {470, 2100}

\bibitem[\protect\citeauthoryear{{Krause} \& {Eifler} et~al.,}{{Krause}
  et~al.}{2017}]{Krause2017}
{Krause} E.,  et~al. 2017, arXiv e-prints, \href
  {https://ui.adsabs.harvard.edu/abs/2017arXiv170609359K} {p. arXiv:1706.09359}

\bibitem[\protect\citeauthoryear{{Landy} \& {Szalay}}{{Landy} \&
  {Szalay}}{1993}]{Landy1993BiasFunctions}
{Landy} S.~D.,  {Szalay} A.~S.,  1993, \mn@doi [\apj] {10.1086/172900}, \href
  {https://ui.adsabs.harvard.edu/\#abs/1993ApJ...412...64L} {412, 64}

\bibitem[\protect\citeauthoryear{{Lee} \& {Huff} et~al.,}{{Lee}
  et~al.}{2019}]{DMASS}
{Lee} S.,  et~al. 2019, \mn@doi [\mnras] {10.1093/mnras/stz2288}, \href
  {https://ui.adsabs.harvard.edu/abs/2019MNRAS.489.2887L} {489, 2887}

\bibitem[\protect\citeauthoryear{{Lee} \& {Troxel} et~al.,}{{Lee}
  et~al.}{2021}]{DMASS-GGL}
{Lee} S.,  et~al. 2021, \mn@doi [arXiv e-prints] {10.1093/mnras/stab3028},
  \href {https://ui.adsabs.harvard.edu/abs/2021arXiv210411319L} {p.
  arXiv:2104.11319}

\bibitem[\protect\citeauthoryear{{Lewis}, {Challinor}  \& {Lasenby}}{{Lewis}
  et~al.}{2000}]{camb}
{Lewis} A.,  {Challinor} A.,   {Lasenby} A.,  2000, \mn@doi [\apj]
  {10.1086/309179}, \href
  {https://ui.adsabs.harvard.edu/abs/2000ApJ...538..473L} {538, 473}

\bibitem[\protect\citeauthoryear{{Limber}}{{Limber}}{1953}]{limber1}
{Limber} D.~N.,  1953, \mn@doi [\apj] {10.1086/145672}, \href
  {https://ui.adsabs.harvard.edu/abs/1953ApJ...117..134L} {117, 134}

\bibitem[\protect\citeauthoryear{{Limber}}{{Limber}}{1954}]{limber2}
{Limber} D.~N.,  1954, \mn@doi [\apj] {10.1086/145870}, \href
  {https://ui.adsabs.harvard.edu/abs/1954ApJ...119..655L} {119, 655}

\bibitem[\protect\citeauthoryear{{LoVerde} \& {Afshordi}}{{LoVerde} \&
  {Afshordi}}{2008}]{LoVerde2008}
{LoVerde} M.,  {Afshordi} N.,  2008, \mn@doi [\prd]
  {10.1103/PhysRevD.78.123506}, \href
  {https://ui.adsabs.harvard.edu/abs/2008PhRvD..78l3506L} {78, 123506}

\bibitem[\protect\citeauthoryear{{Miyatake} \& {More} et~al.,}{{Miyatake}
  et~al.}{2015}]{Miyatake2015}
{Miyatake} H.,  et~al. 2015, \mn@doi [\apj] {10.1088/0004-637X/806/1/1}, \href
  {https://ui.adsabs.harvard.edu/\#abs/2015ApJ...806....1M} {806, 1}

\bibitem[\protect\citeauthoryear{{Moraes} \& {Kneib} et~al.,}{{Moraes}
  et~al.}{2014}]{CFHT-S82}
{Moraes} B.,  et~al. 2014, in Revista Mexicana de Astronomia y Astrofisica
  Conference Series. pp 202--203

\bibitem[\protect\citeauthoryear{OSC}{OSC}{1987}]{OSC}
OSC 1987, Ohio Supercomputer Center, \url {http://osc.edu/ark:/19495/f5s1ph73}

\bibitem[\protect\citeauthoryear{{Okamoto} \& {Hu}}{{Okamoto} \&
  {Hu}}{2003}]{Okamoto2003}
{Okamoto} T.,  {Hu} W.,  2003, \mn@doi [\prd] {10.1103/PhysRevD.67.083002},
  \href {https://ui.adsabs.harvard.edu/abs/2003PhRvD..67h3002O} {67, 083002}

\bibitem[\protect\citeauthoryear{{Padmanabhan} \& {Schlegel}
  et~al.,}{{Padmanabhan} et~al.}{2007}]{Padmanabhan2007}
{Padmanabhan} N.,  et~al. 2007, \mn@doi [\mnras]
  {10.1111/j.1365-2966.2007.11593.x}, \href
  {https://ui.adsabs.harvard.edu/\#abs/2007MNRAS.378..852P} {378, 852}

\bibitem[\protect\citeauthoryear{{Perlmutter} \& {Aldering}
  et~al.,}{{Perlmutter} et~al.}{1999}]{Perlmutter1999}
{Perlmutter} S.,  et~al. 1999, \mn@doi [\apj] {10.1086/307221}, \href
  {https://ui.adsabs.harvard.edu/abs/1999ApJ...517..565P} {517, 565}

\bibitem[\protect\citeauthoryear{{Planck Collaboration} \& {Ade}
  et~al.,}{{Planck Collaboration}}{2016}]{Planck2015CosmologicalParameters}
{Planck Collaboration} 2016, \mn@doi [\aap] {10.1051/0004-6361/201525830},
  \href {https://ui.adsabs.harvard.edu/abs/2016A&A...594A..13P} {594, A13}

\bibitem[\protect\citeauthoryear{{Planck Collaboration} \& {Aghanim}
  et~al.,}{{Planck Collaboration}}{2020a}]{Planck2018CosmologicalParameter}
{Planck Collaboration} 2020a, \mn@doi [\aap] {10.1051/0004-6361/201833910},
  \href {https://ui.adsabs.harvard.edu/abs/2020A&A...641A...6P} {641, A6}

\bibitem[\protect\citeauthoryear{{Planck Collaboration} \& {Aghanim}
  et~al.,}{{Planck Collaboration}}{2020b}]{Planck2018Lensing}
{Planck Collaboration} 2020b, \mn@doi [\aap] {10.1051/0004-6361/201833886},
  \href {https://ui.adsabs.harvard.edu/abs/2020A&A...641A...8P} {641, A8}

\bibitem[\protect\citeauthoryear{{Prat} \& {S{\'a}nchez} et~al.,}{{Prat}
  et~al.}{2018}]{Prat2018DarkLensing}
{Prat} J.,  et~al. 2018, \mn@doi [\prd] {10.1103/PhysRevD.98.042005}, \href
  {https://ui.adsabs.harvard.edu/abs/2018PhRvD..98d2005P} {98, 042005}

\bibitem[\protect\citeauthoryear{{Reid} \& {Ho} et~al.,}{{Reid}
  et~al.}{2016}]{Reid2016}
{Reid} B.,  et~al. 2016, \mn@doi [\mnras] {10.1093/mnras/stv2382}, \href
  {https://ui.adsabs.harvard.edu/\#abs/2016MNRAS.455.1553R} {455, 1553}

\bibitem[\protect\citeauthoryear{{Reyes} \& {Mandelbaum} et~al.,}{{Reyes}
  et~al.}{2010}]{Reyes2010}
{Reyes} R.,  et~al. 2010, \mn@doi [\nat] {10.1038/nature08857}, \href
  {https://ui.adsabs.harvard.edu/abs/2010Natur.464..256R} {464, 256}

\bibitem[\protect\citeauthoryear{{Riess} \& {Filippenko} et~al.,}{{Riess}
  et~al.}{1998}]{Riess1998}
{Riess} A.~G.,  et~al. 1998, \mn@doi [\aj] {10.1086/300499}, \href
  {https://ui.adsabs.harvard.edu/abs/1998AJ....116.1009R} {116, 1009}

\bibitem[\protect\citeauthoryear{{Ross} \& {Samushia} et~al.,}{{Ross}
  et~al.}{2015}]{SDSS-MGS}
{Ross} A.~J.,  et~al. 2015, \mn@doi [\mnras] {10.1093/mnras/stv154}, \href
  {https://ui.adsabs.harvard.edu/abs/2015MNRAS.449..835R} {449, 835}

\bibitem[\protect\citeauthoryear{{Rozo} \& {Rykoff} et~al.,}{{Rozo}
  et~al.}{2016}]{Rozo2016RedMaGiC:Data}
{Rozo} E.,  et~al. 2016, \mn@doi [\mnras] {10.1093/mnras/stw1281}, \href
  {https://ui.adsabs.harvard.edu/\#abs/2016MNRAS.461.1431R} {461, 1431}

\bibitem[\protect\citeauthoryear{{Sachs} \& {Wolfe}}{{Sachs} \&
  {Wolfe}}{1967}]{ISW}
{Sachs} R.~K.,  {Wolfe} A.~M.,  1967, \mn@doi [\apj] {10.1086/148982}, \href
  {https://ui.adsabs.harvard.edu/abs/1967ApJ...147...73S} {147, 73}

\bibitem[\protect\citeauthoryear{{Salazar-Albornoz} \& {S{\'a}nchez}
  et~al.,}{{Salazar-Albornoz} et~al.}{2017}]{Salazar-Albornoz2017}
{Salazar-Albornoz} S.,  et~al. 2017, \mn@doi [\mnras] {10.1093/mnras/stx633},
  \href {https://ui.adsabs.harvard.edu/abs/2017MNRAS.468.2938S} {468, 2938}

\bibitem[\protect\citeauthoryear{{Scolnic} \& {Jones} et~al.,}{{Scolnic}
  et~al.}{2018}]{Scolnic2017}
{Scolnic} D.~M.,  et~al. 2018, \mn@doi [\apj] {10.3847/1538-4357/aab9bb}, \href
  {https://ui.adsabs.harvard.edu/abs/2018ApJ...859..101S} {859, 101}

\bibitem[\protect\citeauthoryear{{Sheldon} \& {Huff}}{{Sheldon} \&
  {Huff}}{2017}]{METACAL1}
{Sheldon} E.~S.,  {Huff} E.~M.,  2017, \mn@doi [\apj]
  {10.3847/1538-4357/aa704b}, \href
  {https://ui.adsabs.harvard.edu/abs/2017ApJ...841...24S} {841, 24}

\bibitem[\protect\citeauthoryear{{Simpson} \& {Heymans} et~al.,}{{Simpson}
  et~al.}{2013}]{Simpson2013}
{Simpson} F.,  et~al. 2013, \mn@doi [\mnras] {10.1093/mnras/sts493}, \href
  {https://ui.adsabs.harvard.edu/abs/2013MNRAS.429.2249S} {429, 2249}

\bibitem[\protect\citeauthoryear{{Singh} \& {Alam} et~al.,}{{Singh}
  et~al.}{2019}]{Singh2019}
{Singh} S.,  et~al. 2019, \mn@doi [\mnras] {10.1093/mnras/sty2681}, \href
  {https://ui.adsabs.harvard.edu/\#abs/2019MNRAS.482..785S} {482, 785}

\bibitem[\protect\citeauthoryear{{Takahashi} \& {Sato} et~al.,}{{Takahashi}
  et~al.}{2012}]{Takahashi2012}
{Takahashi} R.,  et~al. 2012, \mn@doi [\apj] {10.1088/0004-637X/761/2/152},
  \href {https://ui.adsabs.harvard.edu/abs/2012ApJ...761..152T} {761, 152}

\bibitem[\protect\citeauthoryear{{\swap{Torre}{de la }} \& {Jullo}
  et~al.,}{{\swap{Torre}{de la }} et~al.}{2017}]{Torre2017}
{\swap{Torre}{de la }} S.,  et~al. 2017, \mn@doi [\aap]
  {10.1051/0004-6361/201630276}, \href
  {https://ui.adsabs.harvard.edu/abs/2017A&A...608A..44D} {608, A44}

\bibitem[\protect\citeauthoryear{{Tr{\"o}ster} \& {Asgari}
  et~al.,}{{Tr{\"o}ster} et~al.}{2020}]{KiDS1000MG}
{Tr{\"o}ster} T.,  et~al. 2020, arXiv e-prints, \href
  {https://ui.adsabs.harvard.edu/abs/2020arXiv201016416T} {p. arXiv:2010.16416}

\bibitem[\protect\citeauthoryear{{Troxel} \& {MacCrann} et~al.,}{{Troxel}
  et~al.}{2018}]{Troxel2018}
{Troxel} M.~A.,  et~al. 2018, \mn@doi [\prd] {10.1103/PhysRevD.98.043528},
  \href {https://ui.adsabs.harvard.edu/abs/2018PhRvD..98d3528T} {98, 043528}

\bibitem[\protect\citeauthoryear{{Weaverdyck} et~al.}{{Weaverdyck}
  et~al.}{prep}]{WeaverdyckPrep}
{Weaverdyck} N.,  et~al., prep

\bibitem[\protect\citeauthoryear{{Weinberg}}{{Weinberg}}{1989}]{Weinberg1989}
{Weinberg} S.,  1989, \mn@doi [Reviews of Modern Physics]
  {10.1103/RevModPhys.61.1}, \href
  {https://ui.adsabs.harvard.edu/abs/1989RvMP...61....1W} {61, 1}

\bibitem[\protect\citeauthoryear{{Zaldarriaga} \& {Seljak}}{{Zaldarriaga} \&
  {Seljak}}{1998}]{Zaldarriaga1998}
{Zaldarriaga} M.,  {Seljak} U.,  1998, \mn@doi [\prd]
  {10.1103/PhysRevD.58.023003}, \href
  {https://ui.adsabs.harvard.edu/abs/1998PhRvD..58b3003Z} {58, 023003}

\bibitem[\protect\citeauthoryear{{Zhang} \& {Liguori} et~al.,}{{Zhang}
  et~al.}{2007}]{Zhang2007}
{Zhang} P.,  et~al. 2007, \mn@doi [\prl] {10.1103/PhysRevLett.99.141302}, \href
  {https://ui.adsabs.harvard.edu/abs/2007PhRvL..99n1302Z} {99, 141302}

\bibitem[\protect\citeauthoryear{{Zhao} \& {Pogosian} et~al.,}{{Zhao}
  et~al.}{2009}]{mgcamb1}
{Zhao} G.-B.,  et~al. 2009, \mn@doi [\prd] {10.1103/PhysRevD.79.083513}, \href
  {https://ui.adsabs.harvard.edu/abs/2009PhRvD..79h3513Z} {79, 083513}

\bibitem[\protect\citeauthoryear{{Ziour} \& {Hui}}{{Ziour} \&
  {Hui}}{2008}]{Ziour2008}
{Ziour} R.,  {Hui} L.,  2008, \mn@doi [\prd] {10.1103/PhysRevD.78.123517},
  \href {https://ui.adsabs.harvard.edu/abs/2008PhRvD..78l3517Z} {78, 123517}

\bibitem[\protect\citeauthoryear{Zonca \& Singer et~al.,}{Zonca
  et~al.}{2019}]{healpy}
Zonca A.,  et~al. 2019, \mn@doi [Journal of Open Source Software]
  {10.21105/joss.01298}, 4, 1298

\bibitem[\protect\citeauthoryear{{Zucca} \& {Pogosian} et~al.,}{{Zucca}
  et~al.}{2019}]{mgcamb3}
{Zucca} A.,  et~al. 2019, \mn@doi [\jcap] {10.1088/1475-7516/2019/05/001},
  \href {https://ui.adsabs.harvard.edu/abs/2019JCAP...05..001Z} {2019, 001}

\bibitem[\protect\citeauthoryear{{Zuntz} \& {Paterno} et~al.,}{{Zuntz}
  et~al.}{2015}]{COSMOSIS}
{Zuntz} J.,  et~al. 2015, \mn@doi [Astronomy and Computing]
  {10.1016/j.ascom.2015.05.005}, \href
  {https://ui.adsabs.harvard.edu/\#abs/2015A&C....12...45Z} {12, 45}

\bibitem[\protect\citeauthoryear{{Zuntz} \& {Sheldon} et~al.,}{{Zuntz}
  et~al.}{2018}]{Zuntz2018}
{Zuntz} J.,  et~al. 2018, \mn@doi [\mnras] {10.1093/mnras/sty2219}, \href
  {https://ui.adsabs.harvard.edu/abs/2018MNRAS.481.1149Z} {481, 1149}

\makeatother
\end{thebibliography}

 \section*{Affiliations}
 \noindent
{\it 
$^{1}$ Department of Physics, Duke University Durham, NC 27708, USA\\
$^{2}$ Jet Propulsion Laboratory, California Institute of Technology, 4800 Oak Grove Dr., Pasadena, CA 91109, USA\\
$^{3}$ Center for Cosmology and Astro-Particle Physics, The Ohio State University, Columbus, OH 43210, USA\\
$^{4}$ Department of Physics, The Ohio State University, Columbus, OH 43210, USA\\
$^{5}$ Department of Applied Mathematics and Theoretical Physics, University of Cambridge, Cambridge CB3 0WA, UK\\
$^{6}$ Department of Astronomy/Steward Observatory, University of Arizona, 933 North Cherry Avenue, Tucson, AZ 85721-0065, USA\\
$^{7}$ Institute of Physics, Laboratory of Astrophysics, \'Ecole Polytechnique F\'ed\'erale de Lausanne (EPFL), Observatoire de Sauverny, 1290 Versoix, Switzerland\\
$^{8}$ Department of Physics, University of Michigan, Ann Arbor, MI 48109, USA\\
$^{9}$ Institute of Cosmology and Gravitation, University of Portsmouth, Portsmouth, PO1 3FX, UK\\
$^{10}$ Department of Physics, Carnegie Mellon University, Pittsburgh, Pennsylvania 15312, USA\\
$^{11}$ Department of Physics \& Astronomy, University College London, Gower Street, London, WC1E 6BT, UK\\
$^{12}$ Department of Physics and Astronomy, Pevensey Building, University of Sussex, Brighton, BN1 9QH, UK\\
$^{13}$ School of Mathematics, Statistics and Physics, Newcastle University, Newcastle upon Tyne, NE1 7RU, UK\\
$^{14}$ Kavli Institute for Particle Astrophysics \& Cosmology, P. O. Box 2450, Stanford University, Stanford, CA 94305, USA\\
$^{15}$ Kavli Institute for Cosmological Physics, University of Chicago, Chicago, IL 60637, USA\\
$^{16}$ Department of Physics and Astronomy, University of Pennsylvania, Philadelphia, PA 19104, USA\\
$^{17}$ Institute for Astronomy, University of Edinburgh, Edinburgh EH9 3HJ, UK\\
$^{18}$ Jodrell Bank Center for Astrophysics, School of Physics and Astronomy, University of Manchester, Oxford Road, Manchester, M13 9PL, UK\\
$^{19}$ Department of Astronomy, University of California, Berkeley,  501 Campbell Hall, Berkeley, CA 94720, USA\\
$^{20}$ Santa Cruz Institute for Particle Physics, Santa Cruz, CA 95064, USA\\
$^{21}$ Department of Astronomy and Astrophysics, University of Chicago, Chicago, IL 60637, USA\\
$^{22}$ Institut de F\'{\i}sica d'Altes Energies (IFAE), The Barcelona Institute of Science and Technology, Campus UAB, 08193 Bellaterra (Barcelona) Spain\\
$^{23}$ Departamento de F\'isica Matem\'atica, Instituto de F\'isica, Universidade de S\~ao Paulo, CP 66318, S\~ao Paulo, SP, 05314-970, Brazil\\
$^{24}$ Laborat\'orio Interinstitucional de e-Astronomia - LIneA, Rua Gal. Jos\'e Cristino 77, Rio de Janeiro, RJ - 20921-400, Brazil\\
$^{25}$ Fermi National Accelerator Laboratory, P. O. Box 500, Batavia, IL 60510, USA\\
$^{26}$ Instituto de F\'{i}sica Te\'orica, Universidade Estadual Paulista, S\~ao Paulo, Brazil\\
$^{27}$ CNRS, UMR 7095, Institut d'Astrophysique de Paris, F-75014, Paris, France\\
$^{28}$ Sorbonne Universit\'es, UPMC Univ Paris 06, UMR 7095, Institut d'Astrophysique de Paris, F-75014, Paris, France\\
$^{29}$ SLAC National Accelerator Laboratory, Menlo Park, CA 94025, USA\\
$^{30}$ Instituto de Astrofisica de Canarias, E-38205 La Laguna, Tenerife, Spain\\
$^{31}$ Universidad de La Laguna, Dpto. Astrofisica, E-38206 La Laguna, Tenerife, Spain\\
$^{32}$ Center for Astrophysical Surveys, National Center for Supercomputing Applications, 1205 West Clark St., Urbana, IL 61801, USA\\
$^{33}$ Department of Astronomy, University of Illinois at Urbana-Champaign, 1002 W. Green Street, Urbana, IL 61801, USA\\
$^{34}$ Institut d'Estudis Espacials de Catalunya (IEEC), 08034 Barcelona, Spain\\
$^{35}$ Institute of Space Sciences (ICE, CSIC),  Campus UAB, Carrer de Can Magrans, s/n,  08193 Barcelona, Spain\\
$^{36}$ Physics Department, 2320 Chamberlin Hall, University of Wisconsin-Madison, 1150 University Avenue Madison, WI  53706-1390\\
$^{37}$ University of Nottingham, School of Physics and Astronomy, Nottingham NG7 2RD, UK\\
$^{38}$ Astronomy Unit, Department of Physics, University of Trieste, via Tiepolo 11, I-34131 Trieste, Italy\\
$^{39}$ INAF-Osservatorio Astronomico di Trieste, via G. B. Tiepolo 11, I-34143 Trieste, Italy\\
$^{40}$ Institute for Fundamental Physics of the Universe, Via Beirut 2, 34014 Trieste, Italy\\
$^{41}$ Observat\'orio Nacional, Rua Gal. Jos\'e Cristino 77, Rio de Janeiro, RJ - 20921-400, Brazil\\
$^{42}$ Centro de Investigaciones Energ\'eticas, Medioambientales y Tecnol\'ogicas (CIEMAT), Madrid, Spain\\
$^{43}$ Department of Physics, IIT Hyderabad, Kandi, Telangana 502285, India\\
$^{44}$ Faculty of Physics, Ludwig-Maximilians-Universit\"at, Scheinerstr. 1, 81679 Munich, Germany\\
$^{45}$ Department of Astronomy, University of Michigan, Ann Arbor, MI 48109, USA\\
$^{46}$ Institute of Theoretical Astrophysics, University of Oslo. P.O. Box 1029 Blindern, NO-0315 Oslo, Norway\\
$^{47}$ Instituto de Fisica Teorica UAM/CSIC, Universidad Autonoma de Madrid, 28049 Madrid, Spain\\
$^{48}$ Institute of Astronomy, University of Cambridge, Madingley Road, Cambridge CB3 0HA, UK\\
$^{49}$ Kavli Institute for Cosmology, University of Cambridge, Madingley Road, Cambridge CB3 0HA, UK\\
$^{50}$ Department of Physics, Stanford University, 382 Via Pueblo Mall, Stanford, CA 94305, USA\\
$^{51}$ Department of Astronomy, University of Geneva, ch. d'\'Ecogia 16, CH-1290 Versoix, Switzerland\\
$^{52}$ School of Mathematics and Physics, University of Queensland,  Brisbane, QLD 4072, Australia\\
$^{53}$ Max Planck Institute for Extraterrestrial Physics, Giessenbachstrasse, 85748 Garching, Germany\\
$^{54}$ Center for Astrophysics $\vert$ Harvard \& Smithsonian, 60 Garden Street, Cambridge, MA 02138, USA\\
$^{55}$ Australian Astronomical Optics, Macquarie University, North Ryde, NSW 2113, Australia\\
$^{56}$ Lowell Observatory, 1400 Mars Hill Rd, Flagstaff, AZ 86001, USA\\
$^{57}$ George P. and Cynthia Woods Mitchell Institute for Fundamental Physics and Astronomy, and Department of Physics and Astronomy, Texas A\&M University, College Station, TX 77843,  USA\\
$^{58}$ Instituci\'o Catalana de Recerca i Estudis Avan\c{c}ats, E-08010 Barcelona, Spain\\
$^{59}$ Department of Astrophysical Sciences, Princeton University, Peyton Hall, Princeton, NJ 08544, USA\\
$^{60}$ Brookhaven National Laboratory, Bldg 510, Upton, NY 11973, USA\\
$^{61}$ School of Physics and Astronomy, University of Southampton,  Southampton, SO17 1BJ, UK\\
$^{62}$ Computer Science and Mathematics Division, Oak Ridge National Laboratory, Oak Ridge, TN 37831\\
$^{63}$ Universit\"ats-Sternwarte, Fakult\"at f\"ur Physik, Ludwig-Maximilians Universit\"at M\"unchen, Scheinerstr. 1, 81679 M\"unchen, Germany\\
}

\bsp   
\label{lastpage}
\end{document}